\documentclass[sn-basic]{sn-jnl}

\usepackage{graphicx}%
\usepackage{multirow}%
\usepackage{amsmath,amssymb,amsfonts}%
\usepackage{amsthm}%
\usepackage{mathrsfs}%
\usepackage[title]{appendix}%
\usepackage{xcolor}%
\usepackage{textcomp}%
\usepackage{manyfoot}%
\usepackage{booktabs}%
\usepackage{algorithm}%
\usepackage{algorithmicx}%
\usepackage{algpseudocode}%
\usepackage{listings}%

\theoremstyle{thmstyleone}%
\theoremstyle{thmstyletwo}%

\theoremstyle{thmstylethree}%

\begin{document}

\title[Tilted, warped, and eccentric disks]{Tilted, warped, and eccentric disks}


\author[1,2]{\fnm{P. Chris} \sur{Fragile}}\email{fragilep@cofc.edu}
\equalcont{These authors contributed equally to this work.}

\author[3]{\fnm{Adam} \sur{Ingram}}\email{Adam.Ingram@newcastle.ac.uk}
\equalcont{These authors contributed equally to this work.}

\author[4]{\fnm{Gibwa} \sur{Musoke}}\email{gmusoke@cita.utoronto.ca}
\equalcont{These authors contributed equally to this work.}

\author*[5]{\fnm{Gordon I.} \sur{Ogilvie}}\email{gio10@cam.ac.uk}
\equalcont{These authors contributed equally to this work.}

\affil[1]{\orgdiv{Department of Physics \& Astronomy}, \orgname{College of Charleston}, \orgaddress{\street{66 George Street}, \city{Charleston}, \state{SC}, \postcode{29424}, \country{USA}}}

\affil[2]{\orgdiv{Center for Computational Astrophysics}, \orgname{Flatiron Institute}, \orgaddress{\street{162 5th Avenue}, \city{New York}, \state{NY}, \postcode{10010}, \country{USA}}}

\affil[3]{\orgdiv{School of Mathematics, Statistics, and Physics}, \orgname{Newcastle University}, \orgaddress{\city{Newcastle upon Tyne}, \postcode{NE1 7RU}, \country{UK}}}

\affil[4]{\orgdiv{Canadian Institute for Theoretical Astrophysics}, \orgname{University of Toronto}, \orgaddress{\street{60 St George St}, \city{Toronto}, \state{ON}, \postcode{M5S 3H8}, \country{Canada}}}

\affil[5]{\orgdiv{Department of Applied Mathematics and Theoretical Physics}, \orgname{University of Cambridge}, \orgaddress{\street{Wilberforce Road}, \city{Cambridge} \postcode{CB3 0WA}, \country{UK}}}


\abstract{We review some of the interesting consequences that tilts, warps, and eccentricities can introduce into the dynamics, thermodynamics, and observational appearance of accreting systems, with an emphasis on disks around black holes and compact stars. We begin with a review of the two types of precession that are associated with eccentric and tilted orbits in general relativity {and Newtonian gravity}. We then discuss the types of accretion systems that may manifest tilted or eccentric disks. In separate sections we discuss first tilted and then eccentric disks, each section covering relevant and interesting observational, theoretical, and numerical results. Next, we explore potential connections between the phenomenology of quasi-periodic oscillations and {either} tilted {or} eccentric disks. Finally, we present some concluding thoughts and discuss future directions this research might take.}


\maketitle

\section{Introduction}\label{sec:introduction}

The classical theory of accretion disks, developed mainly in the 1970s and greatly stimulated by the work of \cite{Shakura73}, describes a thin disk in which the dominant flow is circular orbital motion in a common plane around a massive central body such as a black hole or a compact star. Over the last five decades, many observational discoveries and theoretical enquiries have stimulated the exploration of alternative accretion-disk geometries, which can now be investigated using three-dimensional numerical simulations. A tilted disk is one in which the orbital motion is inclined with respect to the equator of the central body, and an eccentric disk is one in which the orbital motion is elliptical. The shapes of such disks will generally evolve, on a timescale longer than the orbital period, because of pressure and other collective effects in the disk, as well as precession caused by the gravitational field of the central object or a binary companion, and a tilted disk will generally become warped. The internal dynamics of non-planar or non-circular disks is radically different from that of standard accretion disks. This dynamics and the associated radiative properties have the potential to explain a wide variety of observations.

In the remainder of this introduction, we summarize the precession frequencies of inclined or eccentric orbits due to a variety of gravitational effects and their dependence on the distance from the central object. After considering the various reasons why an accretion disk around a black hole or compact star might be tilted or eccentric, we discuss qualitatively how a fluid disk might respond to differential precession.

{Beyond the introduction, \S\ref{sec:tilted} is dedicated to tilted disks. It is organized into separate subsections on observations (of tilted disks and quasi-periodic oscillations or QPOs), theory (both linear and nonlinear, as well as the Bardeen--Petterson effect) and simulations (of tilted thick and thin disks). We follow a similar approach in \S\ref{sec:eccentric} for eccentric disks, with subsections for observation, theory, and simulations}. In \S\ref{sec:qpo_models} we discuss specific models for QPOs that involve tilted or eccentric disks. We conclude  {in \S\ref{sec:conclusion} with a summary of this review and our thoughts on future prospects}.

\subsection{Orbital precession}

Two of the best known consequences of the general theory of relativity are the \emph{apsidal precession} of eccentric orbits (which provided Einstein with the explanation for the anomalous advance of Mercury's perihelion) and the \emph{nodal precession} of orbits that are inclined with respect to the equatorial plane of a spinning body [known as the Lense--Thirring effect \citep{lense1918influence}]. The angular frequencies $\omega_\text{a}(r)$ and $\omega_\text{n}(r)$ of apsidal and nodal precession of a slightly eccentric or inclined orbit of radius~$r$ around a body of mass~$M$ and spin angular momentum~$S$ are given to a first approximation by
\begin{equation}
  \frac{\omega_\text{a}}{\Omega}\approx\frac{3}{x},\qquad
  \frac{\omega_\text{n}}{\Omega}\approx\frac{2a}{x^{3/2}},
\end{equation}
where $\Omega\approx(GM/r^3)^{1/2}$ is the orbital frequency, while $x=r/(GM/c^2)$ and $a=S/(GM^2/c)$ are the orbital radius and the spin angular momentum in gravitational units. The dependencies $\omega_\text{a}\propto r^{-5/2}$ and $\omega_\text{n}\propto r^{-3}$ imply strong \emph{differential precession} when we consider an extended disk that is either eccentric or inclined. Relativistic apsidal precession is faster than nodal precession, approaching the orbital frequency at the innermost stable circular orbit (ISCO). 

Orbital precession also occurs in Newtonian theory because of departures of the gravitational potential from that of a single point mass. The most important effects arise from inner or outer quadrupole potentials and lead to prograde apsidal precession together with retrograde nodal precession at the same rate ($\omega_\text{a}=-\omega_\text{n}>0$). In a circular binary system with mass ratio $q=M_2/M_1$ and orbital separation $r_\text{b}$, a circumprimary orbit of radius~$r$ experiences
\begin{equation}
  \frac{\omega_\text{a}}{\Omega}=-\frac{\omega_\text{n}}{\Omega}\approx\frac{3q}{4}\left(\frac{r}{r_\text{b}}\right)^3.
\end{equation}
In a typical X-ray binary (XRB), relativistic precession ($\omega\propto r^{-5/2}$ or $r^{-3}$) dominates in the inner disk and Newtonian precession ($\omega\propto r^{3/2}$) in the outer disk; both are strongly differential. Contributions from an inner quadrupole are important in circumbinary disks or around rapidly rotating stars, for which
\begin{equation}
  \frac{\omega_\text{a}}{\Omega}=-\frac{\omega_\text{n}}{\Omega}\approx\frac{3q}{4(1+q)^2}\left(\frac{r_\text{b}}{r}\right)^2
  \qquad\text{or}\qquad
  Q\left(\frac{R}{r}\right)^2,
\end{equation}
respectively, where $Q=k\Omega_*^2R^3/GM$ is the dimensionless quadrupole moment of a star with radius $R$ and spin angular velocity $\Omega_*$ and $k$ is the (dimensionless) apsidal motion constant. In the vicinity of a rapidly rotating neutron star, relativistic apsidal precession dominates, while the Lense--Thirring and quadrupolar contributions to nodal precession can be comparable and of opposite sign.

\subsection{Tilted and eccentric disks}

Having established some of the interesting consequences of tilted and eccentric orbits, we now turn to answer the question of why accretion disks may be tilted or eccentric. We start with the obvious cases: tidal disruption events (TDEs) and active galactic nuclei (AGN) after major mergers \citep{Volonteri_Madau_Quataert2005}. For such events, there is no way for the accretion flow at large scales to know about or be affected by the orientation of the spin axis of the black hole; thus, we would expect the disks in these systems to almost always start out tilted. In fact, in such cases, some form of retrograde accretion, where the angular momentum vector of the accreting gas points opposite the spin axis of the black hole, may be just as common as prograde \citep{King08}. TDEs may also naturally involve eccentric disks since the part of the tidally disrupted star that remains bound to the black hole is initially on a set of orbits with eccentricities only slightly less than unity \citep{Rees1988}. 

Even in stellar-mass accretion systems, misalignment may be common \citep{Fragile01, Maccarone02}, as a consequence of the formation paths of black hole X-ray binaries (BHXRBs). In systems where accretion happens due to mass transfer from one member of the binary to the other, the orientation of the outer accretion disk will usually be determined by the binary orbit. This need not be the same as the orientation of the black hole spin, since its direction is largely set when the black hole forms or becomes part of the system. If the black hole joined the binary through multi-body interactions, such as binary capture or replacement \citep[e.g.][]{AntoniniRodriguezEtAl2018, FragioneKocsis2020}, which may be common for X-ray binaries in globular clusters, then the situation is much like a TDE or merger event in an AGN. There would have been no preexisting symmetry, so the resulting system would nearly always harbor a tilted black hole. Even if the black hole formed through the supernova explosion of a member of a preexisting binary, the black hole could still end up tilted with respect to the binary orbit if the supernova explosion were asymmetric \citep{Jonker_Nelemans2004, Fragos10, Martin_Tout_Pringle2010}, as it appears many are. Misalignments can occur even in an initially aligned system as a result of tilting or warping instabilities, including those due to radiation forces \citep{Pringle96,OgilvieDubus01} and magnetic torques \citep{Lai99}.

In binary systems, both circumprimary and circumbinary disks are known to become eccentric in many circumstances, as revealed by a combination of observations, numerical simulations, and theory (see \S\ref{sec:eccentric}). If the binary itself is eccentric, then some eccentricity is imparted to the disk by secular gravitational interaction (as for example Jupiter and Saturn exchange orbital eccentricity over thousands of orbital periods). Even in a circular binary, eccentricity of the disk can grow as a result of instability, as seen in numerous simulations of circumprimary and circumbinary disks. Perhaps the best understood situation is in circumprimary disks in circular binaries of mass ratio $q \lesssim 0.3-0.4$, where the appearance of superhumps in the superoutbursts of SU~UMa cataclysmic variable stars is explained by the disk becoming eccentric through a mode-coupling process involving the 3:1 resonance and then undergoing prograde apsidal precession (see \S\ref{sec:eccentric}); the same process can be expected in low-mass X-ray binaries (LMXBs), although the observational consequences are different.

Following their formation, there is a tendency for tilted accretion disks to eventually align with their black holes due to their mutual tidal interactions and viscous dissipation \citep{Rees78,King05}. There is also a tendency for eccentric disks to circularize over time, although this is not an inevitable consequence of viscosity \citep{Syer92,Ogilvie01}. Yet, for BHXRBs and TDEs, the alignment and circularization timescales may be comparable to or even longer than their observational lifetimes \citep{Scheuer96}. For AGN, each merger event has the possibility of reorienting the central black hole or its fuel supply, potentially resulting in repeated tilted and eccentric configurations \citep{Kinney00}. Therefore, theoretical arguments at least suggest that tilted and eccentric accretion disks may be quite common.

\subsection{Responses to differential precession}

In the absence of hydrodynamic effects, differential nodal precession acting on a tilted disk would cause an increasing twist of the structure, which is a type of warp involving a smooth misalignment of neighboring orbits. However, a fluid disk responds to being warped by developing an internal torque that transports the misaligned components of angular momentum radially through the disk. This torque or angular-momentum flux can be associated with pressure gradients, magnetic fields, viscous or turbulent stresses, waves, or self-gravity \citep[e.g.][]{Ogilvie13}. Depending on the relationship of the torque to the warp, the disk may be able to respond to differential precession by establishing a smooth warp in which the twisting tendency is resisted by the internal torque. If such a balance can be achieved over the radial extent of the disk, then it may be able to precess coherently, somewhat like a rigid body. The global precession rate will then be a weighted average of the local precession rate (e.g., equation~\ref{eqn:precession} below).
In the case of a very extended tilted disk around a spinning black hole, for example, the global precession rate will be very much smaller than the local rate in the inner part of the disk.

Similarly, in an eccentric disk, differential apsidal precession would lead to a twisting of the family of elliptical orbits and therefore to a bunching {or crossing} of the streamlines that is resisted by hydrodynamic effects. The radial propagation of eccentricity through the disk by means of pressure or other hydrodynamic effects may allow the disk to establish a smooth eccentricity profile with coherent precession.

If the disk cannot supply the internal torque needed to resist the differential precession, or if the relationship of the torque to the geometry of the disk leads to an unstable evolution, then the disk may break into radially distinct sections (see~\S\ref{sec:tearing}).

\section{Tilted disks}\label{sec:tilted}

\subsection{Observations of tilted disks}

{Direct observations of tilted disks are difficult to obtain. It would usually require resolving the accretion system over a wide range of scales. One example where a warp has been measured in a disk is the spiral galaxy NGC 4258, where maser emissions observed by VLBA were used to map out the shape of the disk \citep{Miyoshi95}.}

{More often, the presence of a tilt or warp is inferred}. For BHXRBs, this is usually based on observations of relativistic, bipolar jets that have orientations that appear not to be perpendicular to the binary orbital plane. The underlying assumption is that the orientation of the jet is telling us something about the spin axis of the black hole or the orientation of the inner accretion disk. Since the outer disk will presumably lie in the binary plane, this implies some sort of tilt or warp in the system. Examples of systems where such orientation discrepancies have been claimed include GRO J1655-40 \citep{Orosz97}, XTE J1550-564 \citep{Hannikainen01,Orosz02}, V4641 Sgr \citep{Miller02}, GX 339-4 \citep{Miller09}, and MAXI J1820+070 \citep{Poutanen22}. 

{Another approach to inferring tilted or warped disks is from observations of variability. In the X-ray pulsar Her X-1, the 35-day modulation cycle has been modeled since the early days of accretion disk theory using a precessing, tilted (and warped) disk \citep{Katz73,Gerend76} that is also misaligned with respect to the neutron star's spin axis \citep{Scott00}. Other examples of `superorbital' variability have likewise been attributed to precessing, tilted disks \citep{Kotze12}.} Similar arguments have been applied to a few AGN systems, such as NGC 3079 \citep{Kondratko05}, NGC 1068 \citep{Caproni06}, and NGC 4258 \citep{Caproni07}. Variability in the form of QPOs has also sometimes been attributed to tilts or warps, as we will discuss more in the next subsection.

In the future, more direct evidence for tilted disks may come from polarization measurements. {In cases where the polarization is dominated by emission from or reflection off the disk, the degree of polarization can tightly constrain its angle of inclination \citep{Chandrasekhar60, Schnittman09}. In cases where this disagrees with the binary inclination, this can be a direct indicator for a tilted or warped disk \citep{Cheng16}}.

\subsubsection{Quasi-periodic oscillations}
\label{sec:qpos}

\begin{figure}
\includegraphics[width=\linewidth,trim=0mm 0mm 0mm 0,clip]{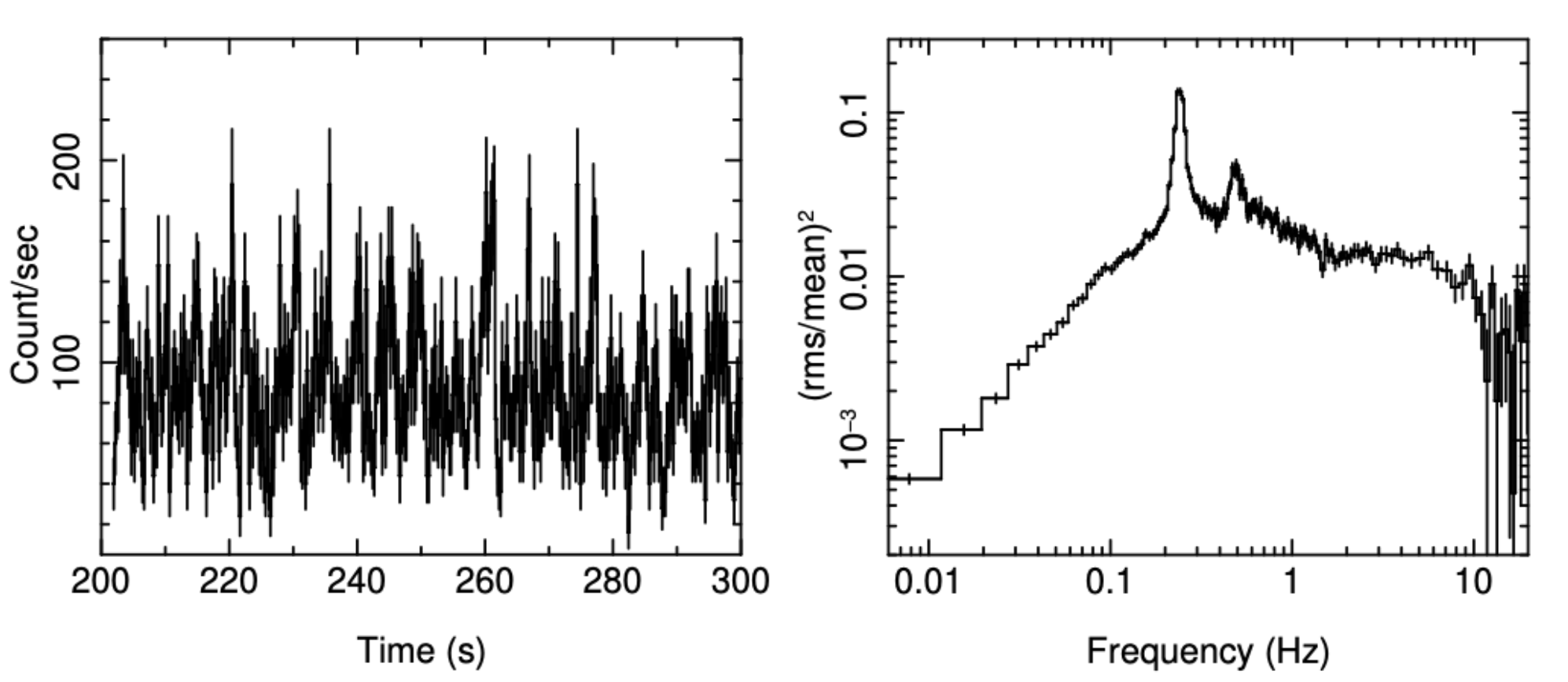}
\caption{{Example of a low frequency QPO observed from the black hole XRB H 1743-322 by \textit{XMM-Newton}. \textit{Left:} A 100 s segment of the $4-10$ keV light curve. \textit{Right:} the power spectrum of the light curve, averaged over a 70 ks interval. The $\sim 4$ s modulation visible in the light curve (left) appears as two narrow, harmonically related peaks in the power spectrum (right). Image reproduced by permission from \cite{Ingram2016} (Figure~2), copyright by RAS.}}
\label{fig:QPO}       
\end{figure}

XRBs generally exhibit temporal variability in their light curves, often in the form of QPOs \citep{Remillard06, Belloni10, Ingram2019}. {Figure~\ref{fig:QPO} shows an example of a prominent QPO observed from the black hole XRB H 1743-322. In the light curve (left), it is possible to see a $\sim 4$ second modulation of the X-ray count rate that is not quite periodic but rather quasi-periodic. This feature can be seen more clearly in the power spectrum, which is the modulus squared of the Fourier transform of the light curve. The $\sim 4$ s flux modulation appears in the power spectrum as a series of harmonically related peaks (in this case only two), with the fundamental peaking at $\sim 0.25$ Hz $=1/(4~{\rm s})$. If the modulation in the flux were perfectly periodic, these peaks would be $\delta-$functions, whereas their finite width signifies that the period and amplitude of the flux modulation vary stochastically in time. The presence of multiple harmonic peaks in the power spectrum signifies that the modulation is non-sinusoidal.}

In black hole systems, {QPOs} are classified into low frequency (LF) if the centroid frequency is $\lesssim 60$ Hz and high frequency (HF) if the centroid frequency is $\gtrsim 60$ Hz. LF QPOs are strong signals, with a fractional rms amplitude as high as $\sim 15\%$ {(as can be seen for the example shown in Figure~\ref{fig:QPO})}. The LF QPO centroid frequency is observed to be tightly correlated with the spectrum of the source, evolving from a minimum of $\sim 0.1$ Hz to a maximum of $\sim 20$ Hz as the peak of the spectrum evolves from hard X-rays to softer X-rays. HF QPOs are much weaker in amplitude and are far more rarely observed \citep{Belloni2012}, but are occasionally observed as a doublet, referred to as the lower and upper HF QPO (with lower and higher centroid frequencies, respectively) \citep{Remillard2002}. 

LF QPOs in neutron star systems behave broadly similarly to their black hole counterparts \citep{vanderKlis2006a}. A common physical origin is thus often assumed, in which case the factor $\sim 5-10$ higher frequencies observed in neutron star sources is the result of the lower mass of the accretor. The neutron star analogues of lower and upper HF QPOs are known as lower and upper kHz QPOs \citep{vanderKlis2006}, with their higher frequencies again consistent with mass scaling. Neutron star kHz QPOs are far stronger and more commonly observed features than black hole HF QPOs. Since kHz QPOs are observed to modulate the boundary layer emission \citep{Gilfanov2003}, it is possible that HF and kHz QPOs are driven by the same underlying mechanism, but kHz QPOs are amplified by the presence of a solid neutron star surface in a way that black hole HF QPOs are not. The centroid frequency of both kHz QPOs is tightly correlated with that of the LF QPO \citep{vanderKlis2006}.

The long association of QPOs with the relativistic frequencies of particle orbits began in the late 1990s with the so-called relativistic precession model (RPM) \citep{Stella1998,Stella1999}, which considers the relativistic frequencies of a test mass at a characteristic radius in the accretion flow. The RPM associates the LF QPO with nodal precession, the lower HF/kHz QPO with apsidal precession, and the upper HF/kHz QPO with the orbital frequency, all at the same characteristic radius. In this picture, the observed increase of the LF and kHz QPO frequencies is caused by the characteristic radius reducing as the source spectrum softens. This simplistic interpretation has yielded some success, most notably for the BHXRB GRO J1655-40, for which there is one observation that simultaneously exhibits a LF QPO and both HF QPOs. Application of the RPM to the three QPO frequencies yields a mass measurement that agrees well with the dynamical mass measurement \citep{Motta2014}, though it requires a rather modest spin.

However, it is unclear how the relativistic frequencies of a test mass at one characteristic radius can modulate the flux. It is possible that over-densities form at some resonant radius in the accretion flow, giving rise to orbiting hot spots, although these hot spots would need to survive for $\sim$hundreds of orbits before being ripped apart by tidal forces to explain the high coherence of some kHz QPOs. More recent efforts to model QPOs have considered how the Lense--Thirring effect influences the dynamics of a tilted and warped disk, which is what we will consider next before eventually returning to the topic of QPO models later in the article (\S\ref{sec:qpo_models}).

\subsection{Theory of tilted and warped disks}
\label{Theory_tilted_warped_disks}

Theoretical treatments of warped disks typically describe the evolving shape of the disk using the tilt vector $\boldsymbol{l}(r,t)$, which is the unit vector that is normal to the local orbital plane at radius $r$ and time $t$. A useful dimensionless measure of the warp amplitude is $|\psi|=|r\,\partial\boldsymbol{l}/{\partial r}|$.
The evolution of mass and angular momentum are described by an extension of the usual conservation equations for an accretion disk \citep{Pringle92}:
\begin{align}
  \frac{\partial\Sigma}{\partial t}+\frac{1}{r}\frac{\partial}{\partial r}(rv_r\Sigma)&=0,\\
  \frac{\partial\boldsymbol{L}}{\partial t}+\frac{1}{r}\frac{\partial}{\partial r}(rv_r\boldsymbol{L})&=\frac{1}{r}\frac{\partial\boldsymbol{G}}{\partial r}+\boldsymbol{T},
\end{align}
where $\Sigma(r,t)$ is the surface density (mass per unit area), $v_r(r,t)$ is the mean radial velocity, $\boldsymbol{G}(r,t)$ is the internal torque (divided by $2\pi$), $\boldsymbol{T}(r,t)$ is the external torque per unit area, and $\boldsymbol{L}=\Sigma r^2\Omega\boldsymbol{l}$ is the angular momentum per unit area, where $\Omega(r)$ is the orbital angular velocity. If the relationship between $\boldsymbol{G}$ (and $\boldsymbol{T}$) and $\boldsymbol{l}$ is known, then these two equations determine $v_r$ and the time-evolution of both the shape ($\boldsymbol{l}$) and the mass distribution ($\Sigma$).

\subsubsection{Linear theory}
\label{sec:linear}

A well-established regime is the linear theory of sufficiently small warps in a laminar, viscous disk \citep[and references therein]{Lubow00}. If the external precession is about the $z$-axis (e.g.\ the spin axis of a central black hole) and the tilts with respect to that axis are small, then the $z$-component of angular momentum is transported as in a flat disk and the accretion process is unaffected by the warp, leading to evolution of $\Sigma$ on the usual viscous timescale. The shape evolves more rapidly, according to
\begin{align}
  \frac{\partial\boldsymbol{l}}{\partial t}&=\frac{1}{\Sigma r^3\Omega}\frac{\partial\boldsymbol{G}}{\partial r}+\omega_\text{n}\boldsymbol{e}_z\times\boldsymbol{l},\label{dldt}\\
  \frac{\partial\boldsymbol{G}}{\partial t}&=\frac{1}{4}\Sigma r^3\Omega^3H^2\frac{\partial\boldsymbol{l}}{\partial r}+\omega_\text{a}\boldsymbol{e}_z\times\boldsymbol{G}-\alpha\Omega\boldsymbol{G},\label{dgdt}
\end{align}
in which only the $x$ and $y$ components of $\boldsymbol{l}$ and $\boldsymbol{G}$ are considered. Here $H$ is the vertical scaleheight of the disk and $\alpha$ is the dimensionless viscosity parameter of \cite{Shakura73}.

The three contributions to $\partial\boldsymbol{G}/\partial t$ in equation~(\ref{dgdt}) are (i) forcing by pressure gradients in the warped disk, (ii) apsidal precession, and (iii) viscous damping. If a flat disk is thought of as consisting of upright columns of gas in hydrostatic equilibrium, then the warp introduces radial pressure gradients that tilt or shear these columns. The internal torque arises from the pressure of columns leaning on their neighbors, although it can also be thought of as due to the radial advection of angular momentum by the shearing flows. Apsidal precession is relevant because the leaning columns can also be interpreted as a stack of elliptical streamlines with an eccentricity proportional to $z$. Precession detunes the resonance that occurs in a Keplerian warped disk in which the epicyclic frequency matches the orbital frequency. Viscous or turbulent damping can also limit the response of the internal torque to the warp; the relevant viscosity parameter $\alpha$ in equation~(\ref{dgdt}) is the one that describes the decay of a vertically shearing flow that oscillates at the orbital frequency.

Even within this linear theory, a number of different phenomena and regimes can be found. Perhaps most important is the distinction between `wavelike' and `diffusive' evolution of the warp in a Keplerian disk \citep{Papaloizou_Lin1995}. If $\alpha\lesssim H/r$ then the warp tends to propagate radially as a bending wave at speed $\Omega H/2$, which is half of the characteristic sound speed for the disk. Such a wave is damped locally at a rate $\alpha\Omega/2$. However, if $\alpha\gtrsim H/r$ then the warp tends to diffuse radially with a diffusivity of $\Omega H^2/(2\alpha)$, which is larger than the effective viscosity $\alpha\Omega H^2$ by a factor $1/(2\alpha^2)$. This `anomalous' behavior can be traced to the resonance in Keplerian warped disks. The rapid evolution of warps in (nearly) Keplerian disks means they that are harder to warp and more likely to respond coherently to differential precession.

The thin, turbulent disks expected in certain states of XRBs and AGN are likely to be in the diffusive regime of warp propagation, while thicker disks or hot accretion flows are likely to be in the wavelike regime. The effect of apsidal precession should also be considered; close to a black hole, the detuning frequency $\omega_\text{a}$ is likely to exceed the damping rate $\alpha\Omega$ of the internal torque.

\subsubsection{The Bardeen--Petterson effect}
\label{sec:bardeen}

Fifty years ago \cite{Bardeen75} posed the question: what is the steady shape of a tilted disk around a spinning black hole? Their answer was a warped shape that makes a smooth transition from to a tilted plane at large radii into the equatorial plane of the black hole at small radii. We refer to this qualititative behavior as BP alignment. 

\begin{figure}
\includegraphics[width=0.32\linewidth,trim=0mm 0mm 0mm 0,clip]{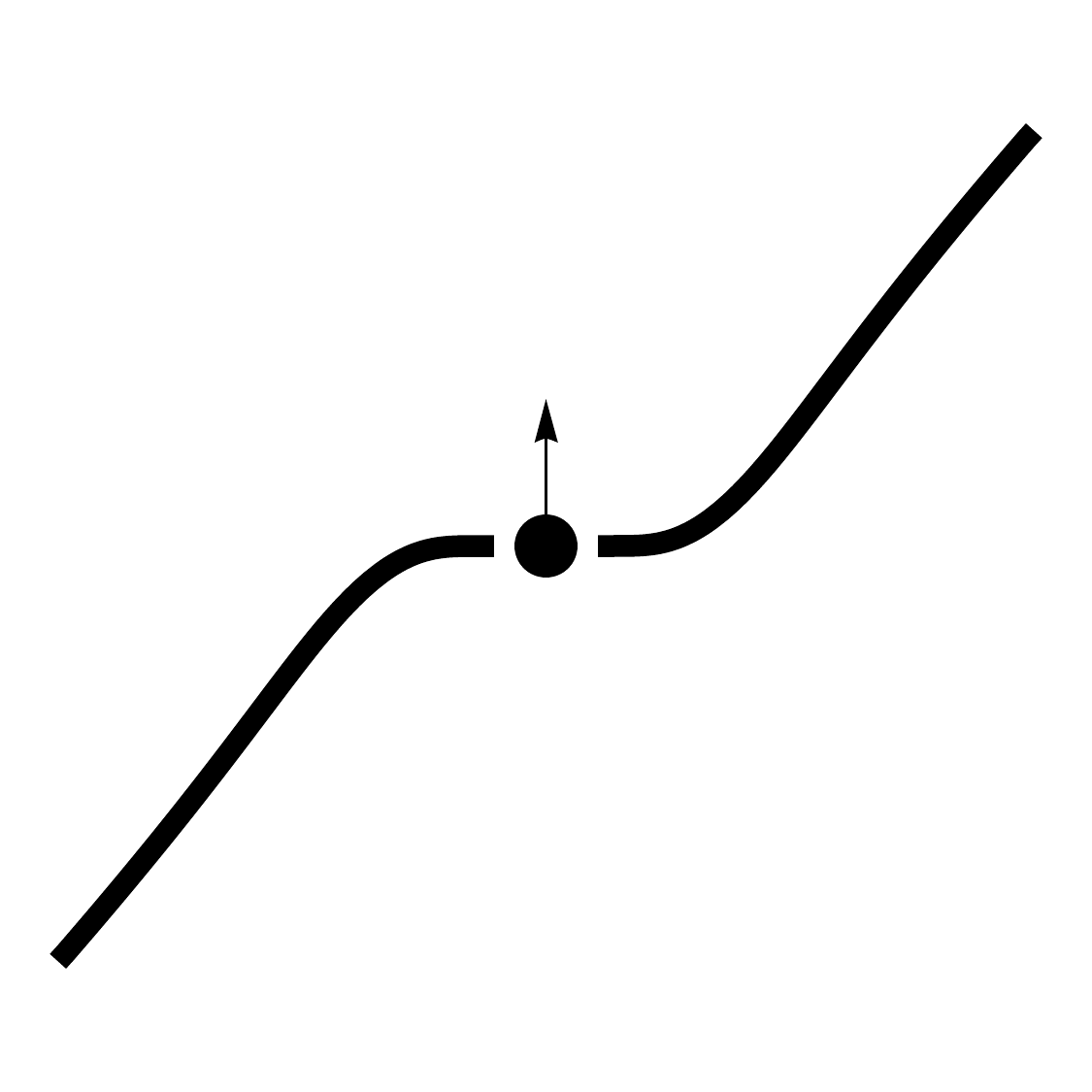}
\includegraphics[width=0.32\linewidth,trim=0mm 0mm 0mm 0,clip]{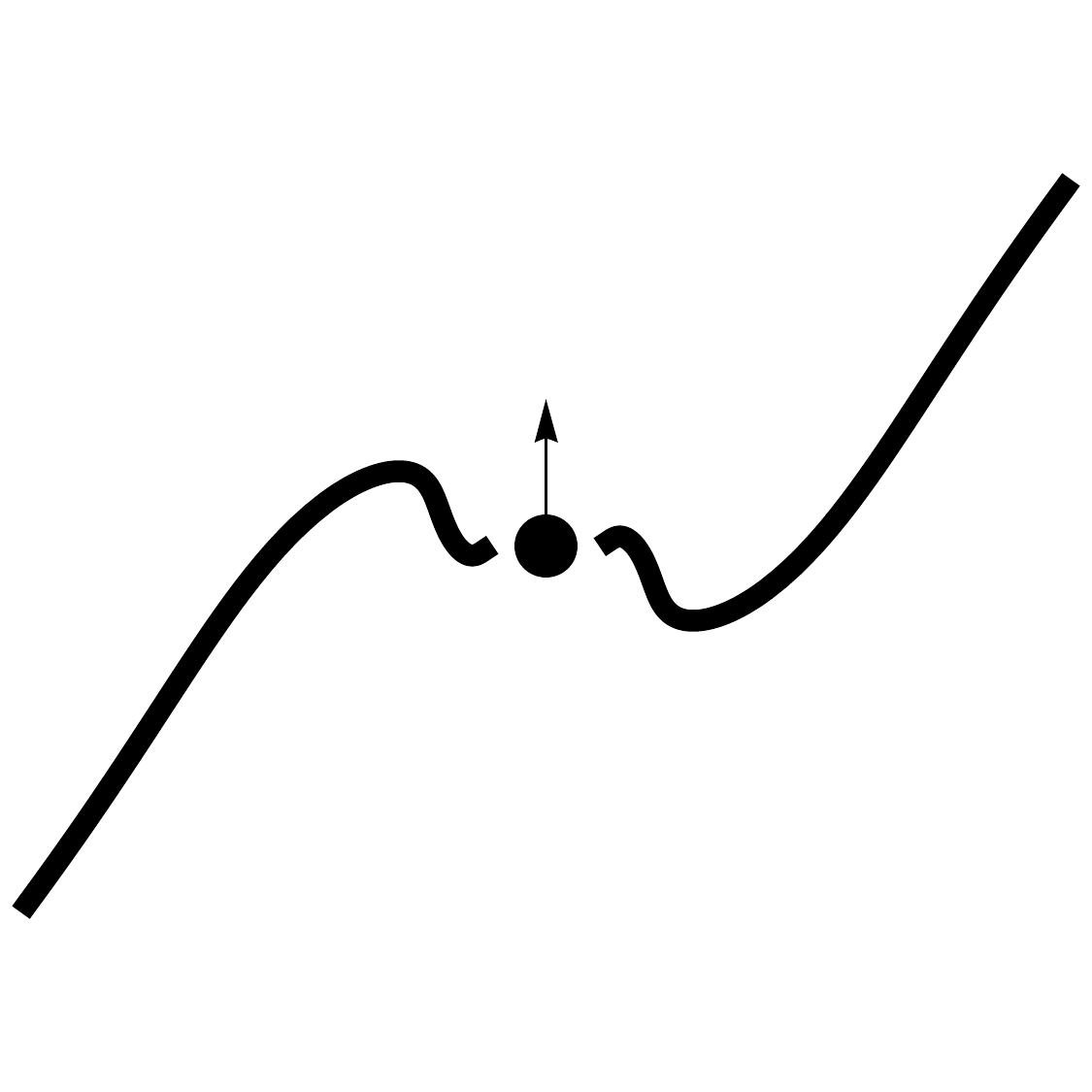}
\includegraphics[width=0.32\linewidth,trim=0mm 0mm 0mm 0,clip]{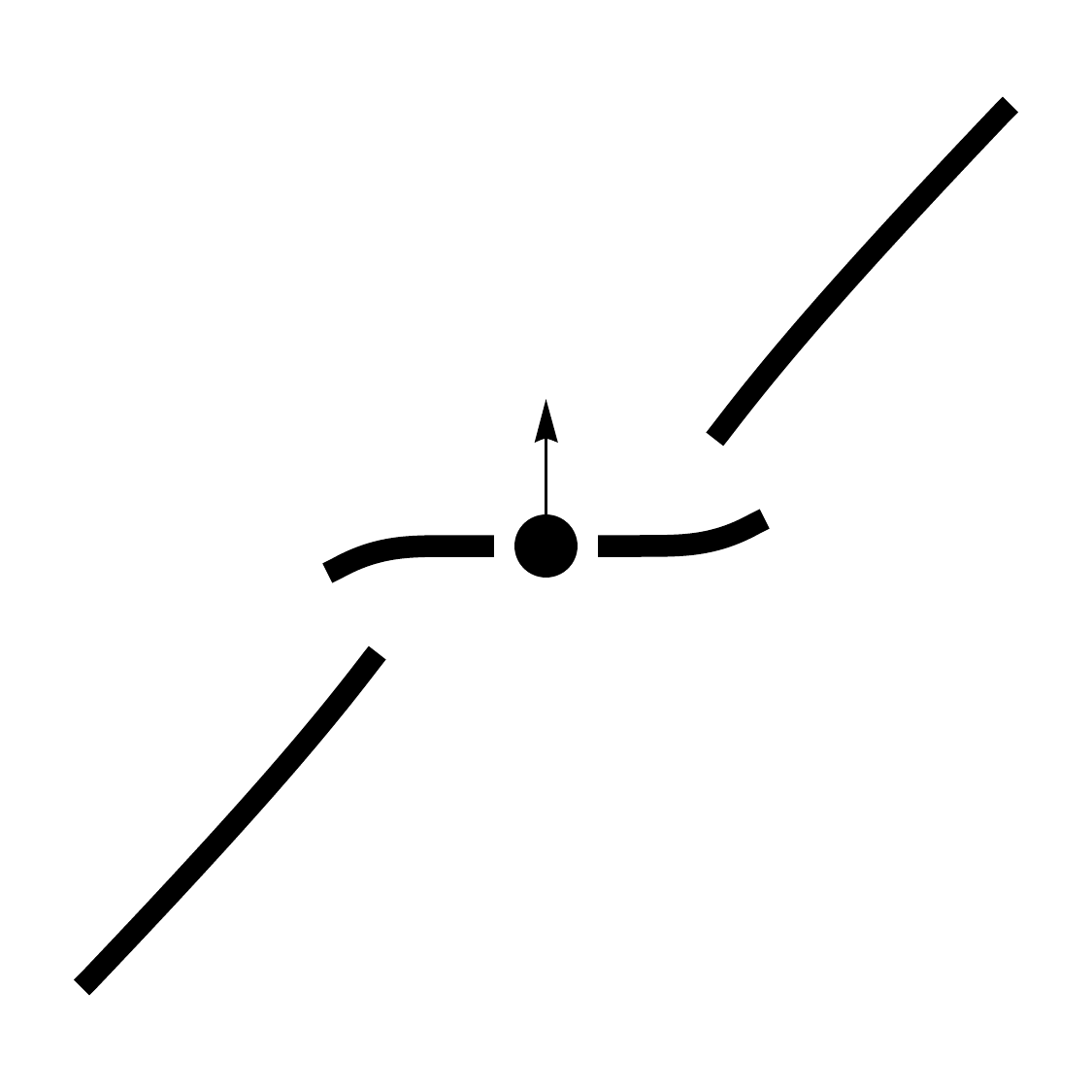}
%
%
\caption{Schematic illustrations of three possible geometries of tilted accretion disks around a spinning black hole. {\em Left:} Smooth, monotonic warp with BP alignment at small radius. {\em Centre:} Smooth, oscillatory warp. {\em Right:} Broken warp with a torn inner disk. The arrows represent the spin axis of the black hole.}
\label{fig:warps}       
\end{figure}

The original analysis \citep{Bardeen75} was flawed as it considered only viscosity and not pressure. Setting the right-hand sides of equations (\ref{dldt}) and (\ref{dgdt}) to zero for a steady warp results in a wave equation for the tilt. When apsidal and nodal precession have the same sign, as for relativistic precession, the local wavelength in units of the scaleheight is \citep{Ivanov97,Lubow02}
\[
  \frac{\lambda}{H}=\frac{\pi\Omega}{\sqrt{\omega_\text{a}\omega_\text{n}}}\approx\frac{\pi x^{5/4}}{\sqrt{6a}}.
\]
This ratio typically exceeds $10$ in the inner disk and increases outwards, implying a large-scale oscillatory warp in the form of a stationary bending wave. Viscosity can damp the wave and limit its amplitude in the inner disk, especially for the thinner disks that occur at lower accretion rates \citep{Ferreira08}, in which case the inner part of the disk does align with the equatorial plane. A sufficiently large viscosity could eradicate the oscillations altogether, resulting in a monotonic warp qualitatively similar to that of \citet{Bardeen75} but with a much smaller transition radius \citep{Nelson00}. Oscillations are also absent if the apsidal precession due to the black hole is neglected. In Figure~\ref{fig:warps} we give schematic illustrations of three possible geometries: BP alignment, an oscillatory warp and a broken warp (see \S\ref{sec:tearing}).

\subsubsection{Nonlinear effects and instability}
\label{sec:nonlinear}

There are several possible nonlinear modifications to the picture presented above \citep{Ogilvie99}. The resonant amplification of the shearing flows inside a warped disk means that nonlinear effects can set in at moderate values of the warp amplitude~$|\psi|$. A nonlinear dependence of the torque $\boldsymbol{G}$ on the warp $\partial\boldsymbol{l}/\partial r$ comes about because of the way the warp and the shear combine to compress the disk twice per orbit. The disk can become significantly non-hydrostatic in the vertical direction, even undergoing a `bouncing' motion in some regimes \citep{Fairbairn21}. The outcome is complicated and depends on the detuning of the Keplerian resonance by apsidal precession or time-dependence of the warp. Typically, the very strong torques and rapid evolution predicted by linear theory in nearly Keplerian disks of low viscosity are moderated in nonlinear theory. An important consequence of this moderation is that the warp evolution can be unstable \citep{Ogilvie00,Dogan18}. While in linear theory the torque $|\boldsymbol{G}|$ is a rapidly (and linearly) increasing function of the warp amplitude $|\psi|$, in nonlinear theory it can reach a maximum at a critical amplitude $|\psi|_\text{crit}$ and decrease thereafter. This leads to an unstable situation in which the steepest part of the warp (with $|\psi|>|\psi|_\text{crit}$) steepens further, leading to the formation of a break in the disk (as well as the breakdown of the theory that predicted it), as described in \S\ref{sec:tearing}.

There are several related manifestations of the nonlinear dependence of the torque on the warp amplitude. When steady states for the radial structure of a laminar, viscous warped disk around a black hole (including nodal but not apsidal precession) are sought, solutions can be found only if the tilt angle is less than a critical value that depends on a combination of parameters including $H/r$, $\alpha$, and $a$ \citep{Tremaine14,Gerosa20}. A related behavior is found for inviscid tilted disks around a Newtonian body with a quadrupole moment \citep{Deng22}. In the first case, as the critical tilt angle is approached, the disk develops a rapid variation in tilt near a critical radius, where the surface density is also depleted; this crisis can be seen as the onset of disk breaking.

Warped disks can also undergo hydrodynamic instability, often known as \emph{parametric instability} because it involves the destabilization of a pair of waves by parametric resonance with the oscillatory motion in the warped disk \citep{Gammie00}. The waves concerned are inertial waves (reliant on the inertial forces associated with orbital rotation and shear), which involve oblique motions with a scale similar to the vertical scaleheight of the disk. Their growth rate is related to the shear rate of the internal flows in the warped disk and so indirectly related to the warp amplitude. The mechanism by which they grow extracts energy from the warp and produces a torque that causes the warp to decay. It is very challenging to capture the parametric instability in global, 3D simulations of warped disks because of the need to resolve waves on scales similar to~$H$ without significant numerical dissipation, though this has been achieved in a few cases \citep[e.g.][]{Deng21}. The parametric instability and the resulting nonlinear development of breaking waves has therefore been studied mainly in local, 2D simulations \citep{Paardekooper19}, which indicate the existence of a turbulent regime in which the torques differ significantly from the laminar theory. The bouncing motions in warped and eccentric disks also give rise to parametric instability \citep{Held24}. 

{In summary, the fluid dynamics of warped disks is potentially very complicated, involving several types of oscillatory motion, instabilities on multiple scales and also shocks \citep{Held26}. Unfortunately, few studies so far have considered the effects of magnetic fields other than providing a turbulent viscosity as a result of the magnetorotational instability (MRI). However, \cite{Paris18} noted that the stiffness imparted by a vertical magnetic field affects the amplitude and profile of shearing motions in a warped disk and therefore the associated internal torques. \cite{FairbairnStone25} have made a first investigation of the interaction between the MRI and the parametric instability in a simplified local model of a warped disk.}

\subsection{Simulations of tilted disks}

A recent review dedicated to tilted disks around black holes, with an emphasis on general relativistic magnetohydrodynamic (GRMHD) simulations, can be found in \cite{FragileLiska25}. In this section, we review some of the most relevant results.

\subsubsection{Thick disks}
\label{sec:precession}

The first general relativistic simulations of tilted disks were performed in the thick-disk (wavelike) regime as this was, by far, the most feasible numerically. The simulations started from rather small (tens of $GM/c^2$), finite reservoirs of gas, usually in the form of a torus orbiting the black hole, first evolving only the hydrodynamic equations \citep{Fragile05} and later adding magnetic fields \citep{Fragile07}. These were largely numerical experiments, as such tilted configurations had not been thoroughly studied previously. 

\begin{figure}
\includegraphics[width=0.5\linewidth,trim=0mm 0mm 0mm 0,clip]{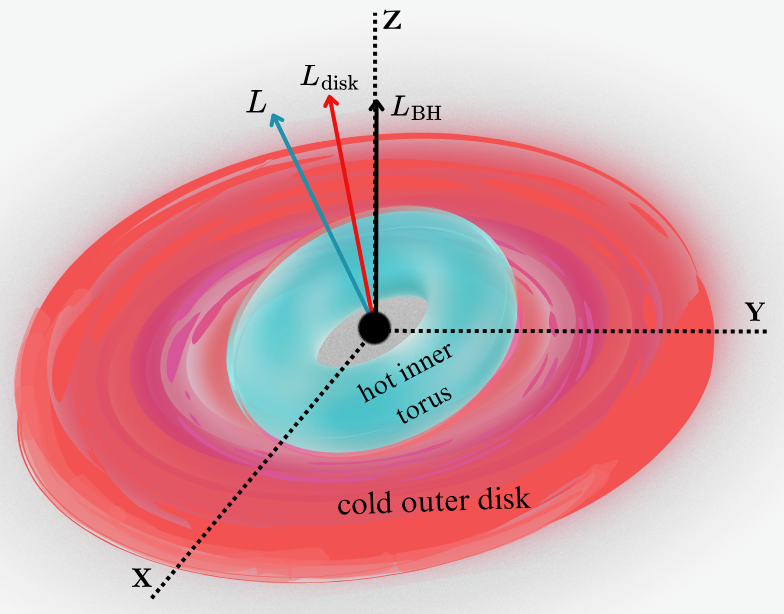}
\includegraphics[width=0.5\linewidth,trim=0mm 0mm 0mm 0,clip]{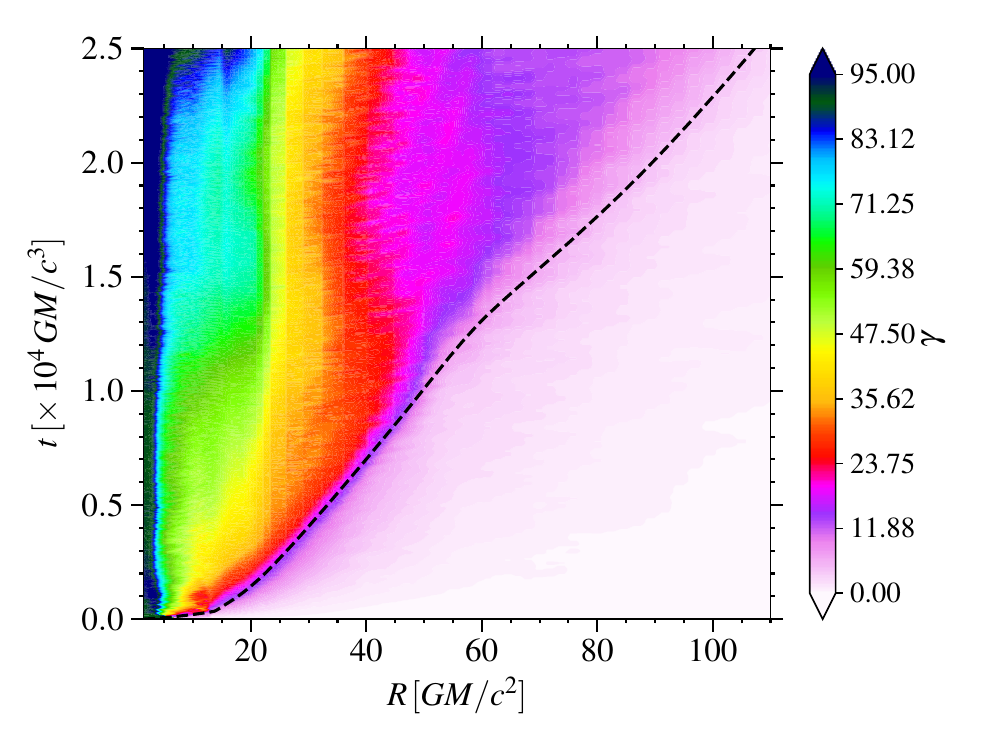}
%
%
\caption{{\em Left:} Geometry of a truncated, tilted disk, with $a = 0.9$, $\beta_0 = 15^\circ$, and a truncation radius of $15\,GM/c^2$. The black-hole $L_\mathrm{BH}$, thin disk $L_\mathrm{disk}$, and torus $L$ angular momenta are indicated. The inner torus precesses about $L_\mathrm{BH}$. {\em Right:} Space-time plot of the cumulative precession angle, $\gamma$ (measured in degrees) as a function of radius. The dashed curve represents the instantaneous position of a bending wave propagating outward through the disk at half the sound speed. The consistent, growing precession angle between 5 and $20\,GM/c^2$ confirms that this region is undergoing rigid-body precession. Images reproduced by permission from \cite{Bollimpalli25} and \cite{Bollimpalli23} (Figure~1), copyright by ESO and RAS.}
\label{fig:thinThick}       
\end{figure}

One of the first important discoveries of these simulations was that, after a brief period of differential twisting, the torus thereafter precessed globally as a single structure \citep{Fragile05}, continuing for multiple periods \citep{Fragile08}. Global precession of tilted disks has now been confirmed in both Newtonian \citep{Larwood96, Nelson00} and GRMHD simulations \citep{Fragile07, Fragile08, Liska18, White19}, in which turbulence occurs because of the MRI. A recent extension of this work demonstrated that global precession is even possible when the torus is surrounded and fed by a thin accretion disk \citep{Bollimpalli23}, as illustrated in Figure \ref{fig:thinThick}.

We can even write down an expression for the rate of disk precession. Generically, the precession period for a solid-body rotator with angular momentum $\boldsymbol{J}$ subject to a torque $\tau$ is simply
$T_{\rm prec} = 2\pi (\sin \beta) (J/\tau)$ \citep{Liu02}, where $\beta$ is the tilt angle between $\boldsymbol{J}$ and the black hole spin axis. The important point about this is that the Lense--Thirring torque gets integrated over the entire disk, so it has a cumulative effect. To illustrate, if we assume a power-law radial dependence to the surface density of the form $\Sigma \propto r^{-\zeta}$ and ignore higher order general relativistic corrections, we can write that
\begin{equation}
T_{\rm prec} = \frac{\pi (1+2\zeta)}{(5-2\zeta)}
\frac{x_o^{5/2-\zeta} x_i^{1/2+\zeta} \left[1-(x_i/x_o)^{5/2-\zeta}
\right]} { a \left[1-(x_i/x_o)^{1/2+\zeta}\right]} \frac{GM}{c^3},
\label{eqn:precession}
\end{equation}
where $x_i$ and $x_o$ are the inner and outer radii of the disk in gravitational units, respectively. This quantitative prediction was first confirmed in the simulations of \cite{Fragile07}. For small disks, or isolated disk components, with $x_o \approx 10$, this precession period falls in the proper range to match the LF QPOs seen in XRBs \citep{Ingram09} (see \S\ref{sec:qpos}).

\subsubsection{Thin disks}

\begin{figure}
\includegraphics[width=1.0\linewidth,trim=0mm 0mm 0mm 0,clip]{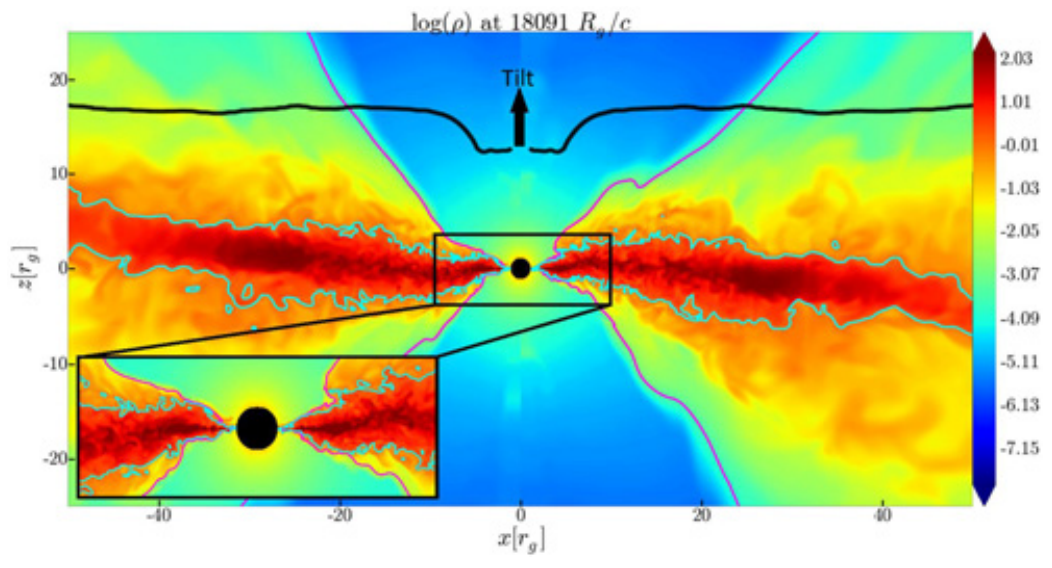}
%
%
\caption{Pseudocolor plot of density for a thin ($H/r = 0.03$), tilted disk, showing that only the innermost part of the disk ($r \lesssim 5\,GM/c^2$) has aligned with the black hole symmetry plane (horizontal in this image). Image reproduced by permission from \cite{Liska19} (Figure~1), copyright by RAS.}
\label{fig:BPLiska}       
\end{figure}

Simulations of tilted, thin disks were first done using smoothed-particle hydrodynamics (SPH) based on a constant, isotropic $\alpha$-viscosity \citep{Nelson00, Lodato07, Lodato10}. Because these simulations shared many of the same assumptions as the leading theoretical models of the time, it was not too surprising that they largely confirmed the predictions in terms of the basic structure and warping radius. However, subsequent GRMHD simulations have shown little to no sign of BP alignment \citep{Teixeira14, Liska19}. Even the thinnest GRMHD simulations ($H/r \approx 0.015-0.03$) show only modest BP alignment out to $\sim 5-10\,GM/c^2$ \citep{Liska19} (see Figure \ref{fig:BPLiska}).

This may be a consequence of the fact that many of the assumptions of the BP model are known to be incorrect. Newtonian gravity plus a torque term is not sufficient to capture all of the relevant relativistic effects \citep{Nelson00}, the disk `viscosity' is not isotropic \citep{Sorathia13, Teixeira14}, nor is it spatially or temporally constant \citep{Penna13}. The implication is that the parameter space associated with the BP effect may be rather small, only including very thin disks. Even then, the alignment may be much more modest than originally predicted.

\subsubsection{Shocks associated with tilted disks}
\label{sec:standing_shocks}

\begin{figure}
\includegraphics[width=0.5\linewidth,trim=0mm 0mm 0mm 0,clip]{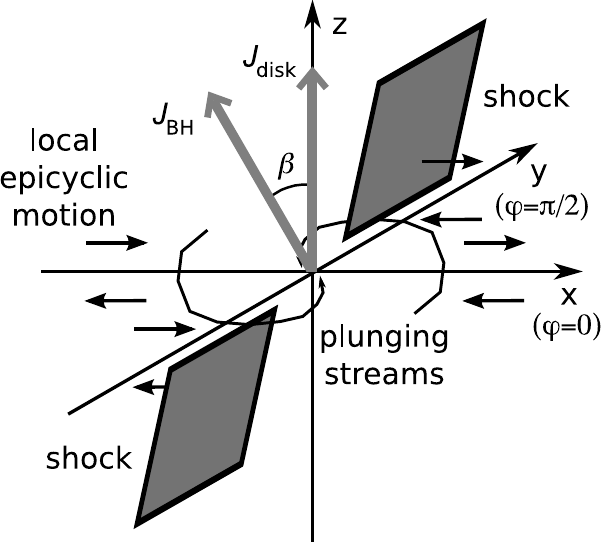}
\includegraphics[width=0.5\linewidth,trim=0mm 0mm 0mm 0,clip]{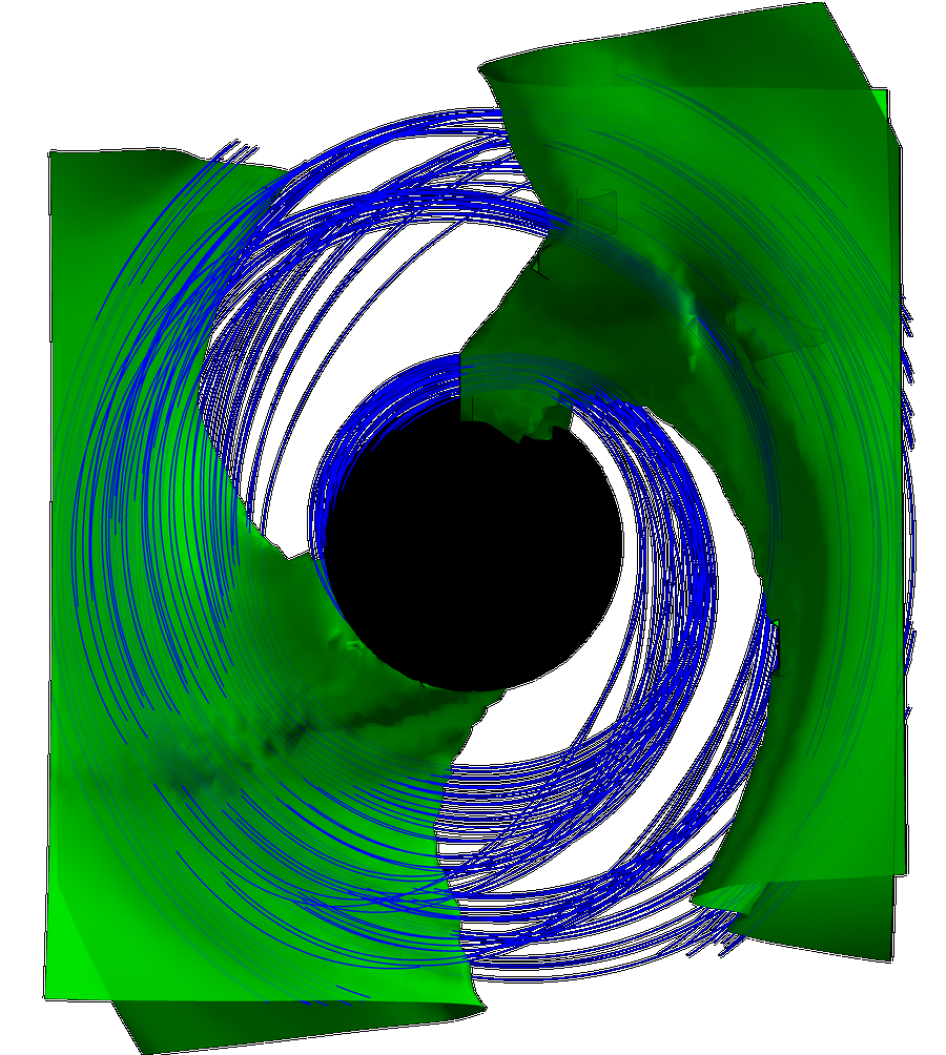}
%
%
\caption{{\em Left:} Schematic diagram of the inner region of the tilted accretion disk, showing the pattern of epicyclic motion, the standing shock, and the plunging streams. {\em Right:} Constant density surface (green) together with selected fluid element trajectories (blue curves) for a $15^\circ$ tilted thick disk simulation. Images reproduced by permission from \cite{Fragile08} (Figure~14) and \cite{Generozov14} (Figure~4), copyright by AAS.}
\label{fig:streams}       
\end{figure}

Tilted thick-disk simulations exhibit interesting non-axisymmetric features, particularly at small radii, that simply are not seen in comparable simulations of untilted disks. For one, there are often found a pair of standing shocks, located on opposite sides of the black hole close to the line of nodes. One is found above the disk midplane and the other below it, as shown in the cartoon diagram in Figure \ref{fig:streams} (left panel). These shocks arise because of an increase in the eccentricity of the disk orbits toward smaller radii \citep{Dexter11}. Thus, the tilt is driving up eccentricity. This phenomenology matches the description in \S\ref{sec:linear}, where the shearing motion inside a warped disk is described as a stack of elliptical streamlines, with eccentricity $\propto z$.
The shocks are a nonlinear manifestation of this motion, in which pressure intervenes to prevent the intersections of streamlines that would otherwise occur.

Associated with the shocks are post-shock density enhancements that appear as arms or bridges of material connecting the disk body with the event horizon (right panel of Figure \ref{fig:streams}). Remarkably these over-dense ridges (and the intervening low-density pockets) are {\em persistent} features, lasting many hundreds of orbital timescales and even precessing along with the rest of the inner disk \citep{Fragile08}. These shocks could potentially be the site of particle acceleration \citep{Sironi24}. They can also extract angular momentum from the regions of the disks closest to the black hole, causing their effective inner radii to be larger than for untilted disks \citep{Fragile09}.

\begin{figure}
\includegraphics[width=1.0\linewidth,trim=0mm 0mm 0mm 0,clip]{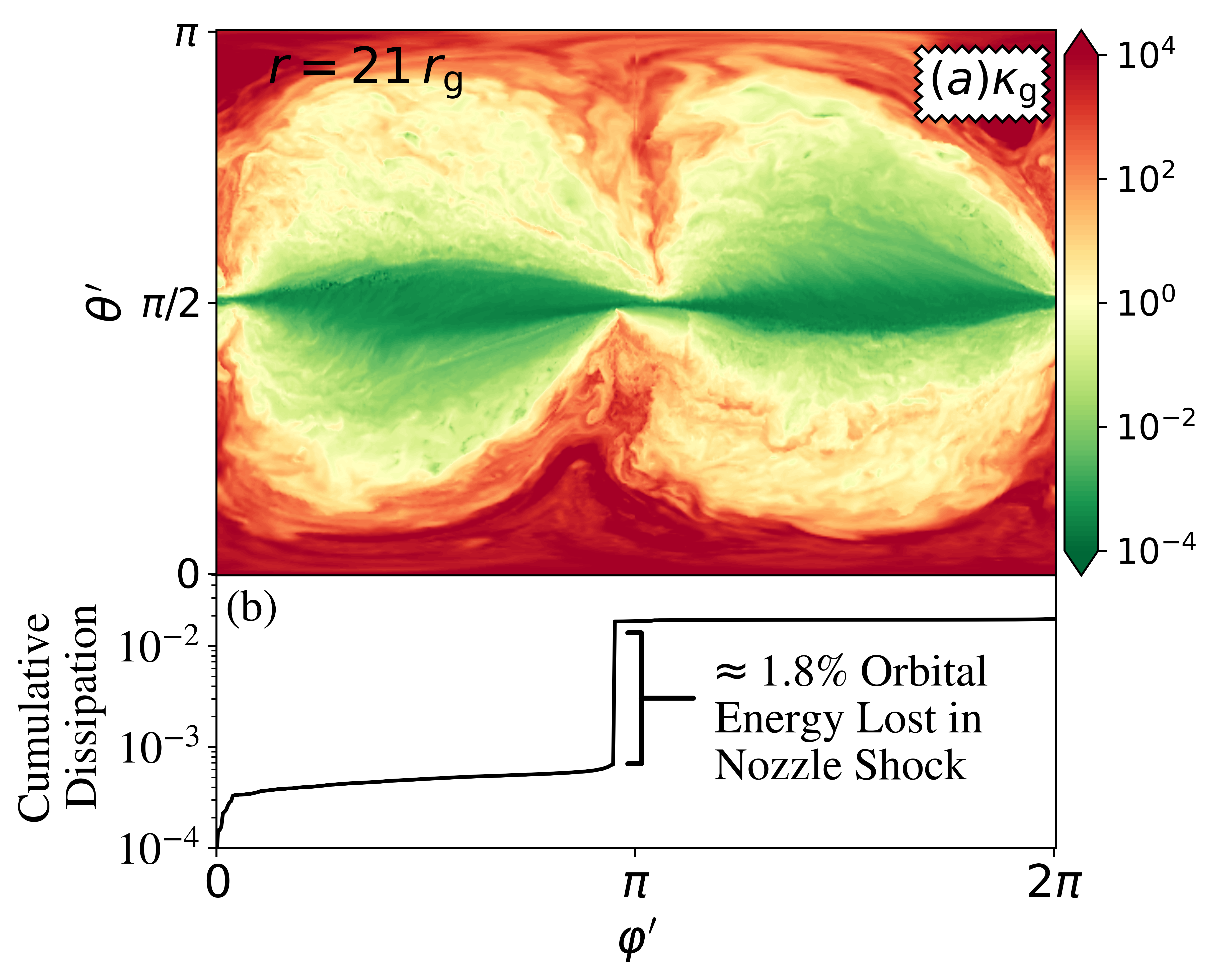}
%
%
\caption{Vertical compression in simulations of warped, thin disks leads to nozzle shocks at two azimuths. Panel (a): Fluid-frame entropy ($\kappa_{\rm g}$) in the $\theta^\prime-\varphi^\prime$ plane showing spikes at $\varphi^\prime\approx0$ (or $2\pi$) and $\pi$, where the disk is most compressed. Panel (b): The cumulative fraction of dissipated energy along the annulus, normalized to the orbital energy. Across the $\varphi^\prime=\pi$ nozzle shock, $\approx1.8\%$ of the orbital energy is dissipated. Image reproduced by permission from \cite{Kaaz23} (Figure~6), copyright by AAS.}
\label{fig:flowNozzleEntropy}
\end{figure}

A different class of shocks are seen in thin disks with large tilts. Here, the disk undergoes vertical compression twice per orbit (see Figure \ref{fig:flowNozzleEntropy} and \S\ref{sec:nonlinear}) that is strong enough to produce shocks, which increase the radial infall speed of the gas by $2-3$ orders of magnitude \citep{Kaaz23}. Whenever a particle passes through one of these `nozzle' shocks (so called by analogy with transonic flows in converging-diverging nozzles), it can lose several percent of its orbital kinetic energy, leading to an effective $\alpha$ of up to $10^2-10^3$.

\subsubsection{Disk tearing in thin, tilted disks}
\label{sec:tearing}

As discussed in \S\ref{Theory_tilted_warped_disks}, whether or not an accretion disk tears or warps depends on whether it is in the wavelike or diffusive warp propagation regime. Disk tearing is expected to occur whenever the Lense--Thirring torque from the warping of spacetime exceeds the viscous torques that hold the disk together \citep[e.g.][]{Dogan18,Nixon_King2012,Raj_Nixon_Dogan2021,Nixon_King_Price2012b,Nealon_Price_Nixon2015}. Generally, if the disk scale height is much smaller than the disk misalignment angle, then viscous diffusion becomes incapable of transporting the rapidly changing angular momentum outwards and the disk tears (see \citealp{Nixon_King2016} for a full review). For a thin, Keplerian, \cite{Shakura73} disk, the radius at which this occurs is estimated to be \citep{Nixon_King_Price_Frank2012}
\begin{equation}
    R_\mathrm{break} \lesssim \bigg( \frac{4}{3} |\sin \beta| \frac{a}{\alpha} \frac{R}{H} \bigg)^{2/3} R_g ~,
\label{eq:breaking_radius}
\end{equation}
where the disk is misaligned by angle $\beta$ with respect to the spin axis of the black hole and $R_g = GM/c^2$ is the black hole gravitational radius. Equation \ref{eq:breaking_radius} implies that systems with thin disks, high black hole spins and large disk misalignment angles will break, since $R_\mathrm{break} > R_g$ in those cases. Furthermore, once the disk starts to tear, it may be the case that the viscosity will drop, promoting further tearing \citep{Ogilvie99,Nixon_King2012,Dogan18}.

\begin{figure}
\includegraphics[width=1.0\linewidth,trim=0mm 0mm 0mm 0,clip]{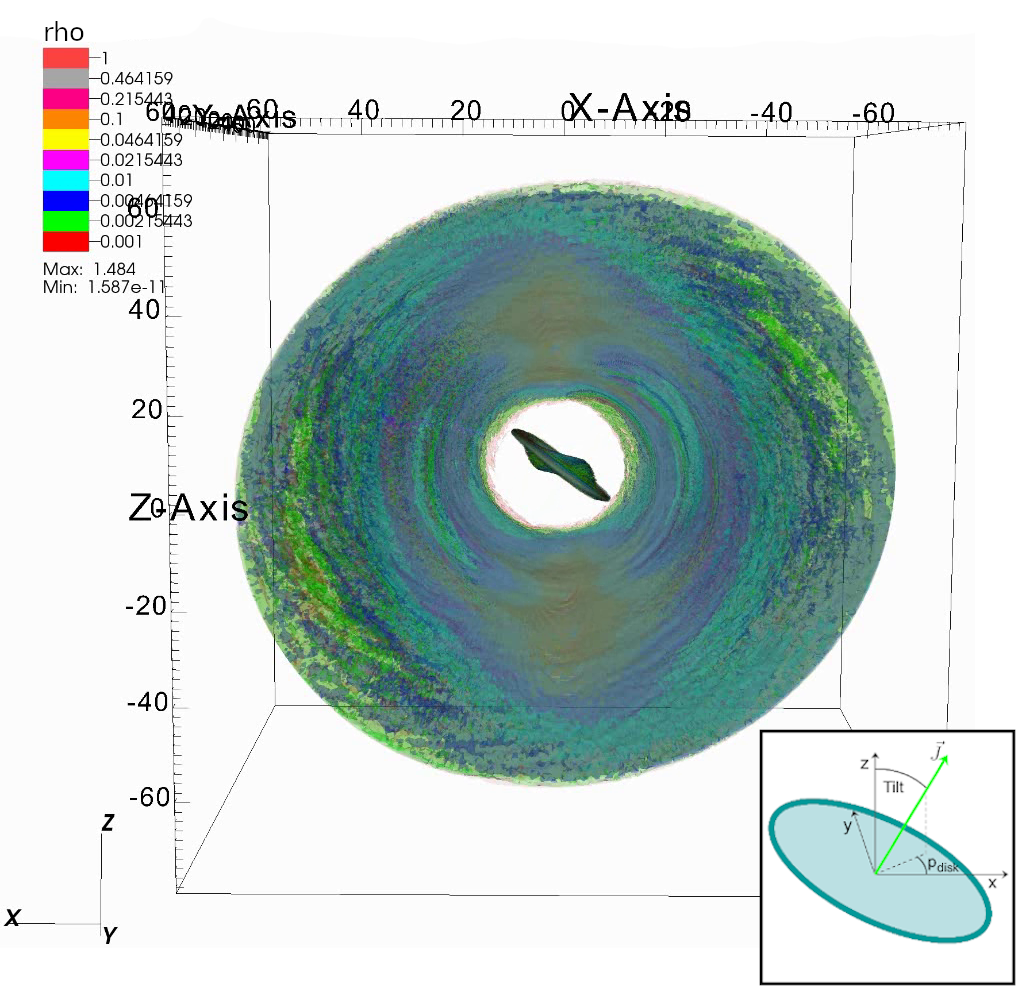}
%
%
\caption{Isocontours of the density of a torn accretion disk. An accretion disk initialized with a high tilt angle ($\beta_0 = 65^\circ$) tears at a radius of $13\,GM/c^2$ into two sub-disks which undergo differential precession. The inner parts
of the sub-disk closest to the black hole undergo BP alignment, which creates a strong warp. Images reproduced by permission from \cite{Musoke23} (Figure~1), copyright by RAS.}
\label{fig:TornDisk}
\end{figure}

Three-dimensional SPH simulations have confirmed that thin, misaligned disks can tear. In fact, they often break into multiple, distinct, narrow rings \citep{Nixon_King_Price2012b,LodatoPringle2006,Nealon_Price_Nixon2015}, with ring thicknesses comparable to the disk scale height $H$. Each ring then precesses at its own rate according to its distance from the black hole. More recent 3D GRMHD simulations \citep[e.g.][]{Liska19,Liska_Bardeen_alignment2021,Kaaz23,Musoke23} likewise verified disk tearing, although with generally wider structures that look more like `sub-disks' than rings (e.g. Figure~\ref{fig:TornDisk}). This difference in tearing behavior between SPH and GRMHD may suggest that magnetic fields or MHD turbulence are somehow affecting the disk tearing process \citep{Liska19}. 

These simulations collectively have shown that disk tearing can result in both disk and jet precession and enhanced jet disruption, dissipation, and misalignment with respect to the spin axis of the black hole. They have also shown that disk tearing can modulate the mass accretion rate through the disk and onto the black hole \citep{Musoke23, Nixon_King_Price_Frank2012, Liska2023}. Furthermore, matter and angular momentum are transferred via streamers of low density gas that connect adjacent sub-disks \citep{Liska19} or rings in the case of SPH simulations \citep{Lodato10}. The streamers can crash into a sub-disk, shocking it and leading to enhanced dissipation \citep{Kaaz23}. Consequently, disk tearing has the potential to explain a wide variety of variability phenomena observed in BHXRBs and AGN, such as flares, disturbed jets, changing-look AGN \citep{Liska2023,Raj_Nixon2021} and QPOs \citep{Fragile01}. 

It should be noted, however, that the large disk tilt angles ($\gtrsim 45^\circ$) typically used in GRMHD simulations of tearing disks are not generally expected to be present in XRBs and AGN. On the other hand, disks that are even thinner than those that are currently computationally feasible (e.g. $H/r < 0.015$), may be able to tear at misalignment angles as small as a few degrees \citep{Nixon_King_Price2012b, Nealon_Price_Nixon2015, Dogan18, Raj_Nixon_Dogan2021}. Therefore, even disks with the moderate misalignment angles expected in XRBs and AGN may be susceptible to disk tearing.

{While simulations of disk tearing are certainly provocative and can plausibly be linked to various observed phenomena, there is currently no convincing observational evidence for torn disks. Furthermore, there are notable inconsistencies in tearing behavior between different numerical approaches (SPH vs. MHD) and computational codes (currently HAMR is the only GRMHD code that has published disk tearing results), and the GRMHD results do not appear to support the predictions of equation~(\ref{eq:breaking_radius}). All of this is to caution that more effort needs to be put into understanding disk tearing.}

\section{Eccentric disks}\label{sec:eccentric}

\subsection{Observations of eccentric disks}

The main observational evidence for eccentric disks around black holes and compact stars comes from \emph{superhumps} in the light curves of some cataclysmic variables (CVs) and LMXBs. Classically, in the SU UMa class of dwarf novae, which typically have mass ratios $q\lesssim 0.3-0.4$, superhumps appear in the optical light curve during a superoutburst, when the accretion disk attains its greatest size, and have a period that is a few percent longer than the orbital period \citep{Patterson05}. Superhumps were first discovered around the time of the work of Shakura \& Sunyaev \citep[see][and references therein]{ODonoghue00} and continue to be studied using \textit{Kepler} and \textit{TESS} \citep[e.g.][]{Bruch23}.

Following computational and theoretical studies in the late 1980s \citep{Whitehurst88}, this phenomenon has been widely related to the outer part of the disk becoming eccentric as a result of an instability associated with the 3:1 resonance. The resonance is between the free eccentric mode, which undergoes slow prograde precession in a non-rotating frame, and a tidal deformation due to the binary companion, which rotates at the binary frequency; the geometry of the disk is therefore modulated at the difference between these two frequencies. The superhumps in SU UMa systems result from the modulation of the viscous dissipation and luminosity of the disk. The elliptical rim of the disk in OY Car was measured directly using eclipses of the hot spot \citep{Hessman92}.

It is natural for LMXBs, especially those with black-hole primaries, to have mass ratios $q\lesssim 0.3-0.4$, and therefore for us to expect superhumps in these systems, too. Indeed, superhumps are seen in some LMXBs, but may be harder to detect: here the optical emission is mainly due to the reprocessing of X-rays from the central source, which is then modulated by the variable disk geometry \citep{Haswell01}.

\subsection{Theory of eccentric disks}

To a first approximation, an eccentric disk can be thought of as made up of {nested} elliptical Keplerian orbits. The evolving shape of an elliptical disk can be described through the dependence of the eccentricity vector $\boldsymbol{e}(r,t)$ on radius\footnote{In nonlinear theory, the non-circular orbits are properly labeled by either their semimajor axis or their semilatus rectum.} $r$ and time $t$. This specifies both the eccentricity $e=|\boldsymbol{e}|$ and the orientation of the ellipse, as the vector~$\boldsymbol{e}$ points towards the periapsis. The variation of $\boldsymbol{e}$ with $r$ cannot be too steep or the streamlines of the disk would intersect their neighbors, leading to shocks.
A typical linear theory for the eccentricity evolution
has the approximate form \citep{Ogilvie08,Teyssandier16}
\begin{equation}
  \frac{\partial\boldsymbol{e}}{\partial t}=\frac{1}{\Sigma r^3\Omega}\frac{\partial}{\partial r}\left[\Sigma r^3\Omega^2H^2\left(\alpha f_1\frac{\partial\boldsymbol{e}}{\partial r}+f_2\,\boldsymbol{e}_z\times\frac{\partial\boldsymbol{e}}{\partial r}\right)\right]+\omega_\text{a}\boldsymbol{e}_z\times\boldsymbol{e},
\end{equation}
where $f_1$ and $f_2$ are positive numbers of order unity depending on the thermal physics. Apart from the apsidal precession $\omega_\text{a}(r)$, the main effect on the shape of the disk is the radial communication of eccentricity by pressure. This is described by the $f_2$ term and corresponds to a dispersive wavelike propagation qualitatively similar to the Schr\"odinger equation. Viscosity (the $f_1$ term) leads to diffusion of the eccentricity on the viscous timescale, a process that is slower than the wave propagation by a factor of order $\alpha$ (although this is a simplification of the effect of viscosity).

In the inner part of a disk around a black hole or neutron star that causes prograde apsidal precession, a balance can be established so that the wave term due to pressure cancels the precessional term at each radius, meaning that the disk does not in fact precess or become twisted. The result is a wave with a wavelength $\lambda$ that depends on radius according to
\begin{equation}
  \frac{\lambda}{H}=2\pi\left(f_2\frac{\Omega}{\omega_\text{a}}\right)^{1/2}\approx2\pi\left(\frac{f_2}{3}\right)^{1/2}x^{1/2},
\end{equation}
where the last approximation is for relativistic precession, in which case $\lambda$ is at least a few times $H$ and increases outwards. This stationary wave pattern is the analogue of the oscillatory warp discussed in \S\ref{sec:bardeen}; the radial variation of the amplitude of the wave is discussed in \cite{Ferreira09}.

The excitation of a global, eccentric mode in a circumstellar disk in a binary system has been treated using the linear theory of eccentric disks \citep{Goodchild06,Lubow10}. As shown by \cite{Lubow91a}, the interaction of an eccentric mode with the tidal deformation in the neighborhood of the $3:1$ resonance causes a localized exponential growth of eccentricity. This process can amplify the mode if the resulting growth rate exceeds the damping rate of the mode through viscosity or other dissipative effects. It has proven difficult to reconcile linear theory, numerical simulations, and observations of superhumps in detail because the $3:1$ resonance is located near the tidal truncation radius for the mass ratios of interest.

The two mechanisms described above can also combine: if the disk develops a global eccentric mode because of the 3:1 resonance in the outermost part of the disk, then that mode effectively produces a stationary pattern of eccentricity in the innermost part of the disk, where the mode's precession rate (itself a small fraction of the binary orbital frequency) is negligible compared to the local angular velocity of the disk. This pattern naturally adopts the wavelike form described above because of relativistic precession. In turn, this wavelike eccentricity gives the inner disk different physical and observational properties and can activate other disk modes as described in \S\ref{sec:nonlinear} and \S\ref{sec:trapped}.

\subsection{Simulations of eccentric disks}

\begin{figure}
\includegraphics[width=1.0\linewidth,trim=0mm 0mm 0mm 0,clip]{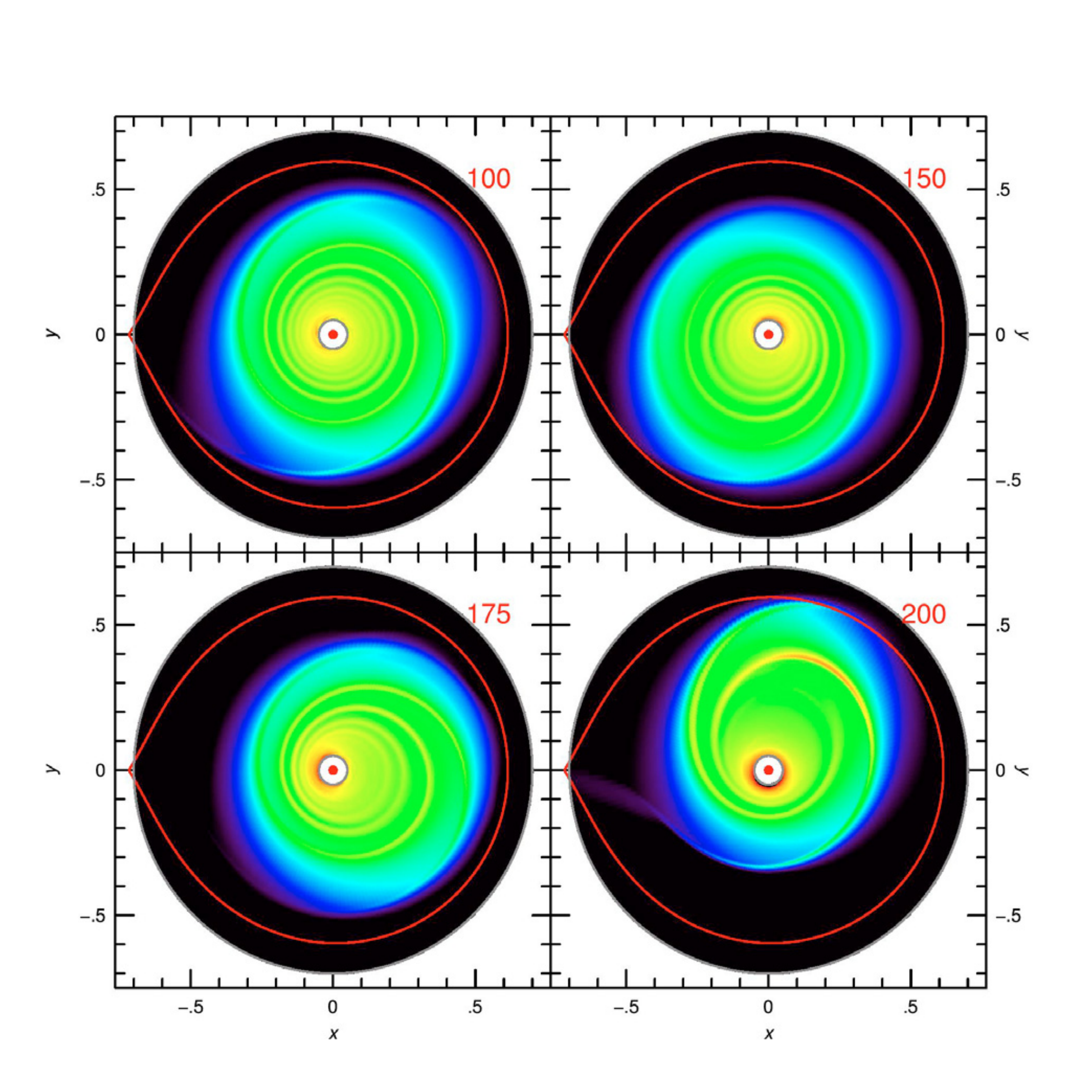}
%
%
\caption{Colormap of the surface density in a 2D grid-based simulation of a precessing eccentric disk around the primary star in a binary system with a circular orbit and a mass ratio of $0.1$. The disk has $H/r=0.05$ and is viewed after 100, 150, 175, and 200 orbital periods. The red curve indicates the Roche lobe of the primary star (central red dot). Image reproduced by permission from \cite{Kley08}, copyright by ESO.}
\label{fig:kley}
\end{figure}

Early simulations of eccentric disks designed to explain superhumps in cataclysmic variables \citep{Whitehurst88,Lubow91b} were carried out using particle-based methods such as SPH. The typical outcome \citep{Whitehurst94,Murray98,Smith07} was an eccentric disk with coherent prograde precession, in good qualitative agreement with observations. For many years it was impossible to replicate these results using grid-based codes. When eccentric disks were finally obtained in grid-based simulations (Figure~\ref{fig:kley}), they were typically much more elliptical than in the SPH simulations \citep{Kley08}. The nonlinear mechanism by which the exponential growth of eccentricity is limited has not yet been clearly identified. The physics is complicated by the fact that the eccentricity is excited in a region close to where the disk is tidally truncated.

\begin{figure}
\includegraphics[width=1.0\linewidth,trim=0mm 0mm 0mm 0,clip]{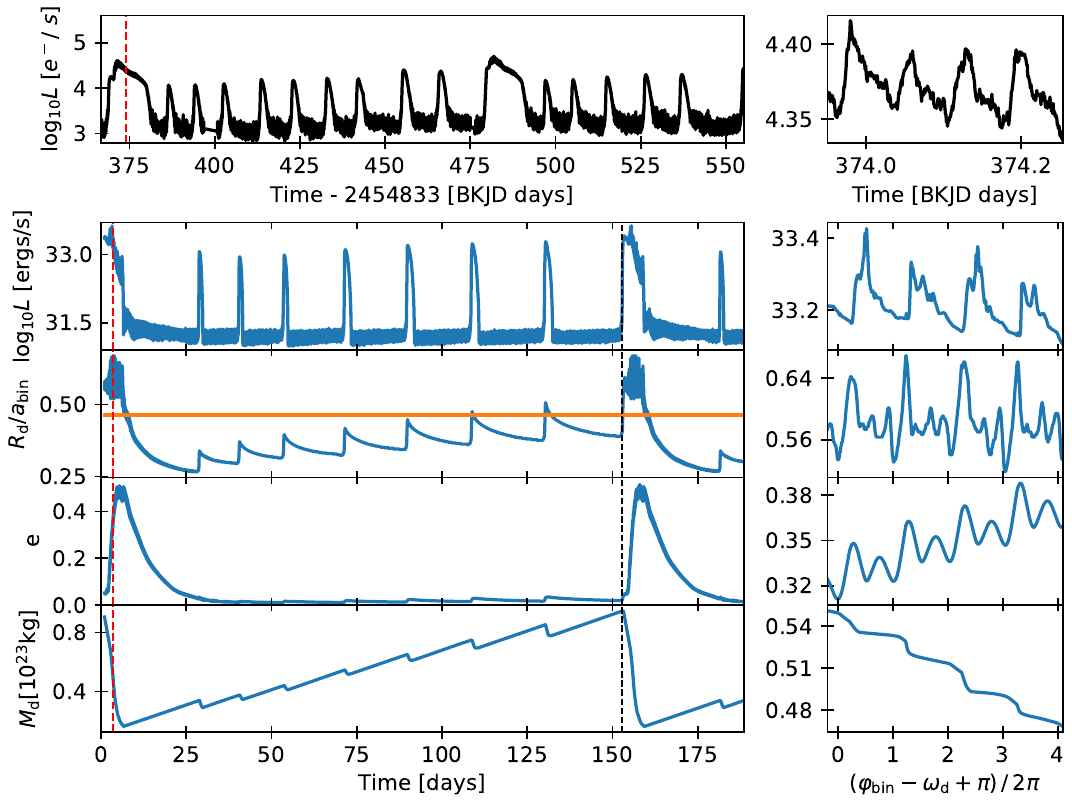}
%
%
\caption{{Top: \emph{Kepler} observations of the luminosity of the SU UMa-type cataclysmic variable V1504 Cygni (binary period $1.7~\text{h}$), with a zoom-in during a superoutburst (top right). Bottom left: Disk luminosity, radius, eccentricity, and total
mass as a function of time in a 2D numerical simulation of an eccentric disk. The orange line indicates the $3:1$ orbital resonance. Bottom right : Zoom-in during a superoutburst as a function of the longitude of pericenter of the disk relative to the position of the binary. Image reproduced by permission from \cite{Jordan24}, copyright by ESO.}}
\label{fig:jordan}
\end{figure}

{\cite{Jordan24} have carried out 2D grid-based simulations of complete cycles of normal outbursts and superoutbursts in SU UMa systems with a viscosity model designed to cause thermal instability (Figure~\ref{fig:jordan}). They have made a detailed comparison of their simulated lightcurves with observational data from \emph{Kepler}, finding generally good agreement but also some discrepancies that might be resolved by a 3D treatment.}

More recently, grid-based MHD simulations have been used to look for the onset of eccentricity in turbulent disks subject to the MRI \citep{Oyang21,Ohana25}. This work indicates that such disks do indeed become eccentric if they spread sufficiently to cover the 3:1 resonance, although this seems to be more difficult in MHD than in viscous hydrodynamics. There is also evidence of `breaking' of the eccentric disk, in the sense that two different eccentric modes develop that are radially separated by a region of circular streamlines.

Most simulations of \emph{circumbinary} disks show that the inner part of the disk becomes eccentric, regardless of whether the binary orbit is circular or eccentric, and for a wide range of computational methods, physical assumptions and parameter regimes [see \cite{Lai23} for a recent review]. The disk either undergoes coherent prograde precession or is locked in phase to the eccentric binary orbit. The former outcome can be understood as the development of an eccentric mode that is trapped in the inner part of the disk by a combination of the apsidal precession due to the binary and the density profile of the truncated disk \citep{Munoz20}; however, the mechanisms by which the eccentricity grows and saturates remain to be fully elucidated. The complex relationship between the shape and the size of the cavity \citep{Thun17} suggests an interplay between the truncation of the disk and the development of the eccentric mode.

\section{QPO models and tilted or eccentric disks}
\label{sec:qpo_models}

\subsection{Precession models}
\label{sec:IDFmodel}

The precession model of Ingram, Done \& Fragile \citep{Ingram09} was designed to explain black hole LF QPOs with Lense--Thirring precession in a more physically realistic scenario than the test mass orbits considered in the RPM. The geometry illustrated in Figure~\ref{fig:thinThick} (left) is assumed, whereby the outer thin disk is truncated and in the binary orbital plane, whereas the inner thick disk, often referred to as the \textit{X-ray corona}, undergoes solid body precession around the black hole spin axis. Such a truncated disk geometry is often invoked to explain the observed X-ray spectrum, whereby the thin disk contributes a thermal spectrum and the corona emits a power-law spectrum due to repeated Compton up-scatterings of disk photons \citep{Eardley1975}. The observed transition of the spectrum from power-law dominated to thermal is then assumed to be caused by the disk truncation radius moving closer to the black hole, which will also speed up precession, consistent with the observed increase of the QPO frequency. At the time the precession model was first proposed, it was merely an assumption that the disk could stay misaligned down to the truncation radius and thus feed misaligned material to the corona, but recent GRMHD simulations provide more solid evidence that the corona can indeed undergo precession inside of a truncated disk \citep{Bollimpalli23}.

Although the model has also been considered in the context of neutron stars \citep{Ingram2010, Fragile20}, most efforts to test it observationally have focused on black hole systems. Frequency resolved spectroscopy indicates that the QPO modulates the power-law coronal emission, not the thermal disk emission \citep{Sobolewska2006}, which is naturally reproduced by a precessing corona. Moreover, the quasi-periodicity of the LF QPOs is predicted within the model due to accretion rate fluctuations propagating through the corona leading to stochastic fluctuation of the solid body precession frequency \citep{Ingram2011}. These rapid fluctuations in the precession frequency are predicted to correlate positively with fluctuations in the flux on the same timescale, as has been observed \citep{Heil2011}. Population studies using the RXTE archive provide another opportunity to diagnose the QPO origin. Several such studies reveal that QPO properties depend on the inclination angle of the binary system \citep{Motta2015,vandenEijnden2017}, which favors models such as precession whereby the geometry of the corona changes with QPO phase.

The current best evidence in favor of the precession model comes from the iron emission line that originates from coronal X-rays reflecting off the thin disk. The precession model predicts the centroid energy of the iron line to vary with QPO phase due to the line rocking from blue- to redshifted as the precessing corona preferentially illuminates approaching followed by receding sides of the thin disk \citep{Ingram2012}. Such a centroid energy modulation has been observed for H 1743-322 \citep{Ingram2016} and GRS 1915+105 \citep{Nathan2022}, albeit with detailed properties that differ from initial expectations.

The newest frontier for testing the precession model is X-ray polarimetry, since the observed X-ray polarization degree and angle are both expected to be modulated on the QPO period if the corona is precessing \citep{Ingram2015, Fragile25}. The first opportunity to conduct this test was presented recently when the Imaging X-ray Polarimetry Explorer (IXPE) observed a strong LF QPO from the black hole XRB Swift J1727.8-1613 \citep{Veledina2023}. No polarization modulation was detected \citep{Zhao2024}, but the upper limits are still compatible with the precession model for a reasonable range of parameter space. Better sensitivity can be achieved for future IXPE observations by employing a longer exposure time, since the Swift J1727.8-1613 observation was rather short due to telemetry restrictions.

{Future timing and polarimetry studies will hopefully continue to probe this QPO model. If the model could be confirmed, it would provide useful constraints of the black hole spin (or neutron star moment of inertia), accretion geometry, and possibly disk viscosity.}

\subsection{Trapped modes}
\label{sec:trapped}

As mentioned in \S\ref{sec:qpos}, HF QPOs from accreting black holes are relatively rare and weak, but they are intriguing in that they appear to be repeatable measures of the dynamics of the innermost parts of the disk, where the effects of relativistic gravity are important and the typical orbital frequencies are hundreds of~Hz.

Perhaps the most important finding from the studies of oscillation modes in relativistic disks around black holes (sometimes known as `diskoseismology') is the existence of hydrodynamic modes that are naturally trapped in the inner part of the disk \citep{Okazaki87}. \emph{Inertial modes} happen internal to the disk and generally involve oblique motion in the meridional plane. The modes usually oscillate at some fraction of the local epicyclic frequency $\kappa$. (In some of the literature, inertial modes are labelled as `g~modes' because, like the g~modes due to buoyancy forces in stars, they occupy a lower-frequency part of the spectrum than the acoustic p~modes.) Moving inwards in a Keplerian disk, $\kappa$ increases steadily, as it equals the orbital frequency $\Omega$; but in a disk around a black hole the epicyclic frequency is less than the orbital frequency (related to the phenomenon of apsidal precession), reaches a maximum $\kappa_\text{max}$ at radius $r_\text{max}$ and decreases to zero at the ISCO. Inertial modes can therefore be radially confined near $r_\text{max}$. The mode that is likely to be most robust and observable is the axisymmetric mode with vertical mode number $n=1$, involving a radial ``sloshing'' motion that periodically compresses the disk along with some corrugation of the midplane. Its frequency is slightly below $\kappa_\text{max}$ by an amount that depends on $H/r$ \citep{Perez97}. The 67~Hz HF QPOs of GRS 1915+105 have previously been associated with this mode \citep{Nowak97}.

Such a trapped mode is likely to be excited at some level by turbulent motion in the disk, although this was not found in MHD simulations \citep{Reynolds09}. The mode may also be damped by turbulent motion. The observed quality factors of HF QPOs are small, suggesting that, if they are trapped modes, the damping is quite severe. It may even be that sufficiently strong toroidal magnetic fields in disks can destroy trapped modes altogether \citep{Fu09}. 

\begin{figure}
\includegraphics[width=1.0\linewidth,trim=0mm 0mm 0mm 0,clip]{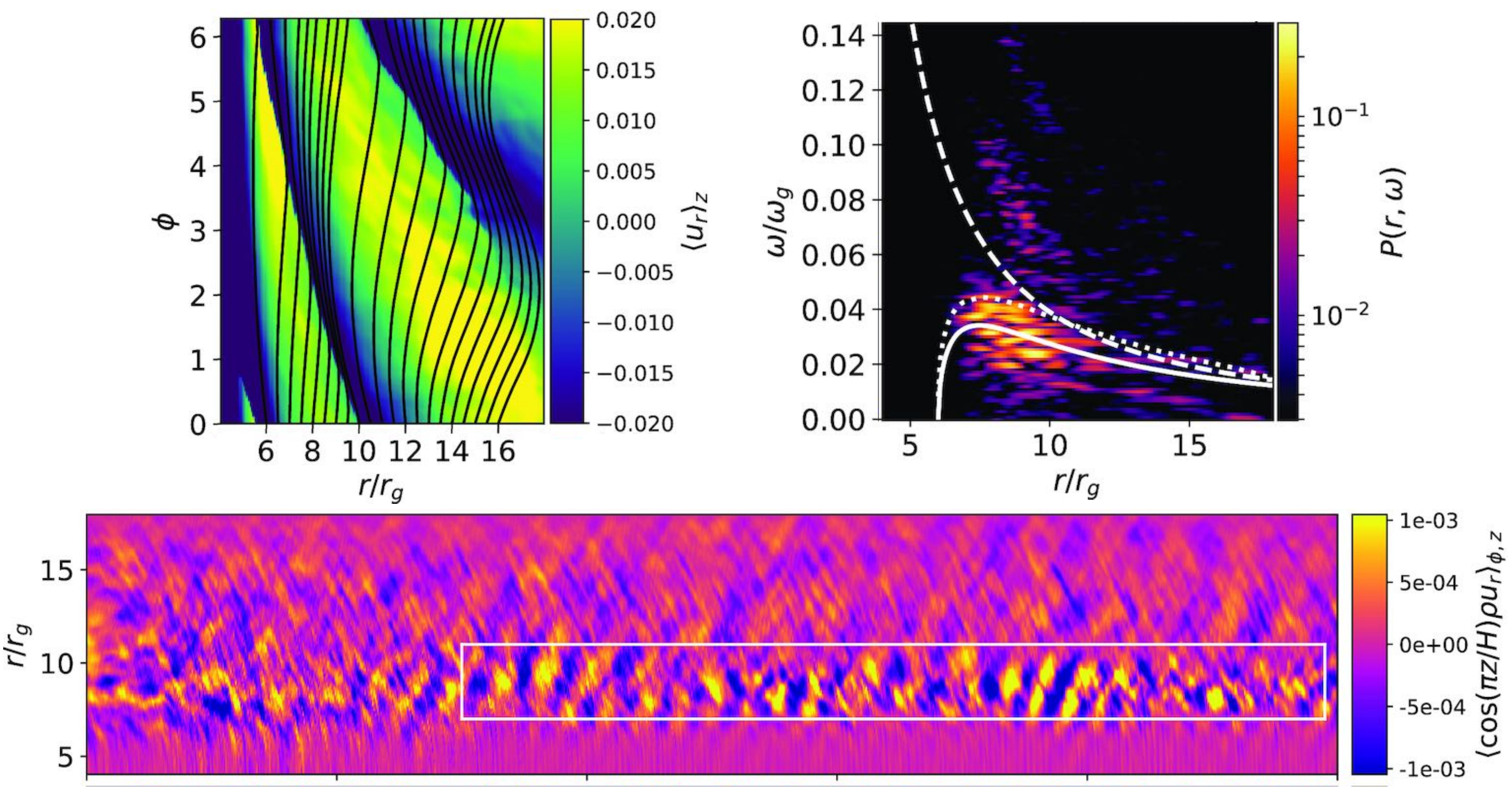}
%
%
\caption{{\em Top left:} Streamlines of the vertically averaged flow in an MHD simulation of the inner part of an eccentric disk around a black hole, represented in Newtonian gravity by using a modified gravitational potential. The large-scale eccentric mode takes a wavelike form in the inner disk and has steepened owing to nonlinearities. {\em Top right:} Power spectrum of radial motions with the vertical structure of the lowest-order inertial mode. The solid and dashed white lines show the epicyclic and orbital frequencies, respectively, while the dotted line shows the epicyclic--Alfv\'enic frequency below which the lowest-order inertial mode is confined. {\em Bottom:} Spacetime diagram of the same radial motions, showing the development of a trapped inertial mode in the inner part of the disk. Images reproduced with permission from \cite{Dewberry20} (Figures 3, 6 and~8), copyright by RAS.}
\label{fig:dewberry}
\end{figure}

Other mechanisms of excitation have been identified that are more promising but may also require special circumstances, which is in tune with the relative rarity of HF QPOs. If the disk has a large-scale warp or eccentricity that appears in the vicinity of $r_\text{max}$, then a mode-coupling process can cause the trapped mode to grow exponentially, at a rate proportional to the square of the warp or eccentricity \citep{Kato04,Ferreira08,Kato08}. In order for the mode (and the HF QPO) to appear, the warp or eccentricity must appear in the innermost part of the disk with sufficient amplitude to overcome the damping of the mode. MHD simulations support this very idea of trapped modes being excited by an eccentricity in the disk (\citealp{Dewberry20}; see Figure~\ref{fig:dewberry}).

The behavior of inclination and eccentricity in gas disks around black holes (\S\ref{sec:linear}) means that they naturally support oscillatory profiles of warp and eccentricity, which may be helpful in exciting trapped modes. However, to relate the amplitude of these waves to the large-scale warp or eccentricity of the disk requires an understanding of their global propagation and attenuation. Most likely, the warp and eccentricity will reach $r_\text{max}$ when the accretion rate is highest, as in the very high state of XRBs \citep{Ferreira09}. {However, more needs to be done to connect HF QPOs with inertial modes.}

\subsection{Disk tearing models}
\label{sec:QPO_disk_tearing}

Numerical simulations are increasingly supporting the view that LF QPOs may be produced by geometric models -- models in which a constant, intrinsic X-ray flux is modulated by changes in the apparent geometry of the accretion flow in a quasi-periodic manner. The precession model of \citet{Ingram09}, outlined in \S\ref{sec:IDFmodel}, comprises one of the most popular such models. Several GRMHD simulations of finite (relatively small outer radii), geometrically thick, tilted disks have now demonstrated that Lense--Thirring precession is indeed a viable LF QPO mechanism \citep{Fragile05,Fragile07,Fragile08,White19,Bollimpalli23}. However, none of those works directly addressed the question of {\em why} there was a relatively compact, tilted region of the disk that was free to precess in the first place; they simply assumed it was. One possible answer to that question is disk tearing. Any disk, as long as it is sufficiently tilted and thin enough, can tear (see \S\ref{sec:tearing}). Provided it tears at an appropriate radius, this model can produce the required precession frequencies without invoking any additional physics \citep{Liska19,Liska_Bardeen_alignment2021,Kaaz23,Musoke23}. 

As an additional bonus, these tearing disk simulations also show evidence for HF QPOs, which has largely been missing from earlier simulations. For example, the simulation of \cite{Liska_HAMR_2022} produced a pair of HF QPOs with a frequency ratio of 1:2 \citep{Musoke23}, consistent with the HF QPO ratio observed in GRS 1915+105 \citep{Belloni_Altamirano2013}. The HF QPOs corresponded to the local radial epicyclic frequency (and its harmonic) at the disk tearing radius \citep{Musoke23}. A more recent radiative GRMHD simulation of a similar setup has largely confirmed this behavior \citep{Liska2023}.

In a similar vein, GRMHD simulations of truncated, tilted disks have also yielded HF QPO-like features \citep{Bollimpalli24}, corresponding in this case to a vertical oscillation of the entire region inside the truncation radius. This region also produces a LF QPO through Lense--Thirring precession \citep{Bollimpalli23}. 

Though observations are currently unable to constrain either an intrinsic or geometric origin for HF QPOs \citep[e.g.][]{Ingram2019}, the GRMHD simulation results are beginning to suggest that tilted disks show more high-frequency (QPO-like) variability than comparable untilted simulations \citep[][and references therein]{Bollimpalli23}. This implies that disk tilt may be a crucial factor for producing HF QPOs.

{While tilted GRMHD simulations are starting to show interesting QPO-like behavior, more work needs to be done to demonstrate that these oscillations appear in corresponding synthetic light curves with the right variability pattern and power spectra expected from observations. This would also allow easier identification of the corresponding QPO type (see \S \ref{sec:qpos}). However, it remains challenging to run numerical simulations long enough with frequent enough data dumps to produce power spectra that cover the full frequency range of interest (ideally mHz to kHz). Nevertheless, we expect this to be an area of high activity over the next few years.}

\section{Conclusion}\label{sec:conclusion}

In this chapter we have reviewed theoretical, numerical, and observational aspects of accretion disks that are tilted, warped, or eccentric, with an emphasis on disks around black holes and compact stars. These alternatives to the standard accretion disk geometry have been widely explored over the last five decades and are playing an increasing role in explaining accretion phenomena.

We summarize our review of these topics as follows:
\begin{enumerate}

\item Misalignments between the orbital angular momentum of accreting matter and either the spin angular momentum of the central object or the orbital angular momentum of a binary companion are commonplace. Tilted disks naturally become warped because the nodal precession rate (due to relativistic or Newtonian gravitational effects) depends strongly on the radius of the orbit. Warping is resisted by internal torques associated mainly with pressure and shearing motions.

\item A laminar, viscous warped disk has a complicated nonlinear behavior that can include a tendency for the warp to steepen into a discontinuous break. Such behavior can be seen in numerical simulations of tilted disks subject to differential precession, using a variety of numerical methods, including those that resolve MRI turbulence.

\item Depending on the physical parameters, a tilted disk around a spinning black hole may not undergo a smooth transition into the equatorial plane at small radii; it may instead adopt an oscillatory profile or it may break or tear into independently precessing rings or sub-disks. Standing shocks can also be expected in the accretion flow.

\item Eccentric disks commonly occur in and around binary systems and after TDEs, and they can be long-lived. Owing to relativistic precession, a large-scale eccentric mode can naturally present a wavelike form in the vicinity of a black hole.

\item Warped and eccentric disks have a rich internal dynamics because of the oscillatory geometry experienced by orbital fluid elements. Rather than being in hydrostatic balance, the disk undergoes systematic oscillations which can destabilize secondary waves and oscillation modes.

\item These dynamical phenomena have the potential to explain a wide range of observations of accreting black holes and neutron stars. In particular, a promising explanation of (type-C) LF QPOs is the nodal precession of a tilted inner component of the accretion flow, which could be either a hot inner flow (in the case of a truncated disk) or a ring or sub-disk that has torn from a tilted disk. Furthermore, HF QPOs could be explained by radial or vertical oscillations of such a torn ring or relativistically trapped inertial modes activated by a warp or eccentricity.

\end{enumerate}

{Finally, we leave the reader with some open questions to ponder regarding tilted, warped, and eccentric disks:}
\begin{enumerate}

\item {Is it possible to observationally confirm the existence of tilted or eccentric disks? Since resolving accretion disks on the scale of black holes or compact stars is generally not possible, our best hope likely will either come from high time-resolution X-ray spectroscopy (resolving accretion processes in the time rather than space domain) or X-ray polarimetry.} 

\item {Is it possible to definitively tie any QPOs to precession, tearing, or inertial modes associated with tilt or eccentricity? Again, we may have to turn to time-resolved spectroscopy or polarimetry.}

\item {Do the QPOs found in fluid variables from numerical simulations of tilted disks, such as velocity or mass accretion rate, manifest themselves in synthetic light curves? If so, do they have the same properties as observed QPOs? These questions should be answerable, in principle, from existing simulations, although this will likely require careful modeling of the radiation.}

\end{enumerate}

\backmatter


\bmhead{Acknowledgements}
We are grateful to Alexandra Veledina and the conveners of the workshop on \textit{Accretion Disks: The First 50 Years}, generously hosted by the International Space Science Institute, Bern, that led to the writing of this review.
PCF gratefully acknowledges the support of NASA under award No 80NSSC24K0900. The Flatiron Institute is a division of the Simons Foundation. AI acknowledges support from the Royal Society. GM is supported by a Canadian Institute of Theoretical Astrophysics (CITA) post-doctoral fellowship and acknowledges support from the Simons Collaboration on Extreme Electrodynamics of Compact Sources (SCEECS). GIO acknowledges the support of the Science and Technology Facilities Council (STFC) through grant ST/X001113/1. 

\section*{Declarations}

The authors have no competing interests to declare that are relevant to the content of this article.

\bibliography{ms}

\newpage

\begin{appendices}

\end{appendices}

\end{document}